\documentclass[12pt,reqno]{amsart}

\usepackage[T1]{fontenc}
\usepackage{textcomp}
\usepackage[cp1252]{inputenc}
\usepackage{tikz-cd}
\usepackage{amsthm}
\usepackage{graphicx} 
\DeclareUnicodeCharacter{00A0}{~}        
\DeclareUnicodeCharacter{2013}{-}       
\DeclareUnicodeCharacter{2014}{-}      
\DeclareUnicodeCharacter{2212}{-}        
\DeclareUnicodeCharacter{2018}{'}        
\DeclareUnicodeCharacter{2019}{'}        
\DeclareUnicodeCharacter{201C}{``}       
\DeclareUnicodeCharacter{201D}{''}       
\DeclareUnicodeCharacter{2026}{\ldots}   

\usepackage[a4paper,margin=1in]{geometry}
\usepackage{setspace}
\usepackage{newtxtext,newtxmath}
\usepackage[final]{microtype}
\usepackage{bbm}

\usepackage{mathtools,amsfonts}
\numberwithin{equation}{section}
\usepackage{amsaddr}

\usepackage{amsthm}

\newtheoremstyle{clay}
  {6pt}{6pt}{\itshape}{}{\bfseries}{.}{0.5em}
  {\thmname{#1}\ \thmnumber{#2}\ (\thmnote{#3})}
\theoremstyle{clay}

\theoremstyle{definition}

\theoremstyle{remark}

\usepackage{enumitem}
\setlist[enumerate,1]{label=(\roman*)}
\usepackage{booktabs}
\usepackage{caption}
\usepackage{braket}
\usepackage{float}
\usepackage{tabularx}
\usepackage{siunitx}

\usepackage{hyperref}
\hypersetup{
  colorlinks=true,
  linkcolor=teal!60!black,
  citecolor=teal!60!black,
  urlcolor=teal!60!black,
  pdftitle={Reflection-Positive Transfer Matrices and a Strong-Coupling Gap for Lattice SU(N) Yang-Mills},
  pdfauthor={Mir Faizal and Arshid Shabir},
  pdfkeywords={Yang-Mills, mass gap, reflection positivity, transfer matrix, lattice gauge theory}
}

\usepackage{bm} 

\usepackage[nameinlink,capitalize,noabbrev]{cleveref}

\usepackage{physics}

\newcolumntype{Y}{>{\centering\arraybackslash}X}
\usepackage{placeins}   
\usepackage{needspace}  
\usepackage[numbers,sort&compress]{natbib}
\begin{document}
\title[Quantum Speed Limit as a Sensitive Probe of Planck-Scale Effects]%
{Quantum Speed Limit as a Sensitive Probe of Planck-Scale Effects}
\author{Salman Sajad Wani \qquad Saif Al-Kuwari}
\address{Qatar Center for Quantum Computing, College of Science and Engineering, Hamad Bin Khalifa University, Doha, Qatar}

\begin{abstract}
Many quantum-gravity scenarios predict a minute modification of the canonical commutator, known as the generalized uncertainty principle (GUP), whose low-energy signatures are, in principle, accessible to state-of-the-art laboratory tests. We compute first-order minimal-length corrections to the quantum speed limit (QSL)  for three cases: uniform superpositions in an infinite square well, coherent harmonic-oscillator states, and squeezed-oscillator states.  We identify a universal amplification law: for any pure state, the fractional shift of either speed limit scales linearly with $\beta$ and algebraically with the state's effective Hilbert-space size. As the effective Hilbert-space dimension can be exceedingly large, the associated minimal-length signatures are amplified by several orders of magnitude. Using high-precision matter-wave timing data, we set a direct bound on the GUP parameter $\beta$, which quantifies minimal-length quantum-gravity effects. Our analysis indicates that phase-locked, short-time overlap fits on kilogram-scale optical-spring modes can tighten this bound by orders of magnitude. We outline two implementable measurement pipelines-continuous back-action-evading single-quadrature readout and stroboscopic, phase-locked pulsed tomography-that exploit this leverage, making QSL-based timing a practical, near-term probe of minimal-length physics on quantum-optical and optomechanical platforms.
\end{abstract}
\maketitle
\section{Introduction}
Several quantum gravity theories predict a fundamental minimal length scale, below which the very concept of distance becomes ambiguous. This idea first arose from studies of string scattering at Planck-scale energies. Attempting to probe distances smaller than the fundamental string length \( l_s \) inevitably excites extended string modes, making it impossible to achieve finer spatial resolution~\cite{GUPAmati1987,GUPGrossMende1988}. Thought experiments combining quantum mechanics and black hole physics strongly support this idea, clearly indicating minimal length as a key signature of Planck-scale physics~\cite{GUPGaray1995}. At low energies, we commonly describe these minimal-length effects using the generalized uncertainty principle (GUP):
$
[\hat{x}, \hat{p}] = i\hbar\left(1 + \beta \hat{p}^{2}\right) + \mathcal{O}(\beta^{2})
$
where the parameter \(\beta\) measures the strength of quantum-gravity corrections~\cite{GUPScardigli1999}. Micro-black-hole thought experiments suggest possible values for \(\beta\), while operator-based approaches directly build these higher-order corrections into quantum Hamiltonians~\cite{GUPDasVagenas2008}.

Beyond GUP-based searches, earlier quantum-gravity phenomenology developed complementary frameworks and observational tests that set stringent bounds on Planck-suppressed effects across multiple sectors. On the theory side, two systematic parameterizations of departures from exact Lorentz symmetry were developed. Doubly special relativity deforms relativistic kinematics by introducing an invariant high-energy scale while retaining observer independence, which modifies the transformation rules and can imprint on high-energy and long-baseline observations \cite{AmelinoCamelia2002Nature}. However, DSR faces several consistency issues, including the ``soccer-ball" problem for composite systems, observer/locality tensions (e.g., relativity of locality), ordering ambiguities in $\kappa$-Minkowski realizations, and unresolved coproduct and multiparticle-sector definitions \cite{Hossenfelder2010PRL,Hossenfelder2014SIGMA,ArzanoMarciano2007PRD,MeljanacKresicJuric2008JPA,AmelinoCamelia2011PRLRelativeLocality,GUPHossenfelder2013}. In parallel, effective field-theory extensions of the Standard Model, organized as the Standard-Model Extension, catalog symmetry-breaking operators by mass dimension and link operator coefficients to specific low-energy observables and Lorentz-violating couplings \cite{Colladay1997PRD,Colladay1998PRD}. On the observational side, high-energy astrophysics led early tests; time-of-flight analyses of gamma-ray bursts and active galactic nuclei use cosmological baselines to probe tiny energy-dependent photon delays, reaching sensitivities at linear Planck suppression (and beyond) \cite{AmelinoCamelia1998Nature,Ellis2000ApJ}. Polarization-based probes searched for birefringence and helicity-dependent dispersion, and synchrotron constraints informed by Crab-Nebula observations added complementary sensitivity to small differential phase accumulations and helicity splittings \cite{GleiserKozameh2001PRD,Jacobson2003Nature}. Within the same phenomenological program, loop-quantum-gravity-motivated dispersion relations were analyzed for fermions, photons, and ultra-high-energy cosmic rays, translating microscopic discreteness or polymeric kinematics into constraints on modified propagation \cite{Alfaro2002PRD,Alfaro2003PRD}. An important theoretical insight concerned radiative stability: generic Lorentz-violating EFTs generate large loop-induced contributions to lower-dimension operators unless protected by symmetry or fine tuning, which reshapes what is technically natural under renormalization and restricts viable deformations \cite{Collins2004PRL}.
Alongside astrophysical baselines and EFT systematics, precision laboratory resonators provide an independent, controlled setting to test minimal-length and GUP-motivated effects through amplitude-frequency relations in macroscopic systems. Sapphire split-bar experiments set the first limits on minimal-length-induced frequency shifts. Cryogenic quartz bulk-acoustic-wave (BAW) resonators then strengthened those bounds by orders of magnitude, and the best BAW constraints are tighter than the sapphire split-bar limits \cite{Bushev2019PRD,Bourhill2015PRApplied,Campbell2023PRD,Bawaj2015NatCommun}. Not all of these bounds are specific to GUP models. They belong to a broader quantum-gravity phenomenology that seeks Planck-suppressed signatures through propagation tests, polarization and dispersion measurements, threshold phenomena, and precision spectroscopy, each with its own systematics and interpretation.

Many current experimental proposals target static effects, including small shifts in energy levels and slight changes in tunnelling rates~\cite{Ali2011,GUPPikovski2012,Gao2017,Bosso2017,Ciszak2024}. These static measurements usually produce very small fractional shifts, which limits sensitivity and makes direct detection difficult. The literature quantifies GUP corrections for wells and barriers~\cite{GUPApplicationQuantumWells2014,GUPApplicationBladoTunneling2016,GUPApplicationGuo2016}, solvable momentum-space models~\cite{GUPApplicationSamarTkachuk2016}, harmonic oscillators~\cite{GUPApplicationChang2002,GUPApplicationPark2020,GUPApplicationDossa2021}, and related scattering and nuclear settings~\cite{GUPApplicationVahedi2012,GUPApplicationHassanabadi2015,GUPApplicationFaruque2014,GUPApplicationMoniruzzaman2018}. This body of work motivates laboratory probes using spectroscopy, tunnelling, and coherence~\cite{Ali2011,GUPHossenfelder2013,GUPPetruzziello2021}. Recently, strategies have shifted to dynamical effects, including decoherence from quantum-gravity corrections~\cite{GUPPetruzziello2021}, modified quantum noise in optomechanical systems~\cite{GUPPikovski2012,Sen2022}, and macroscopic harmonic oscillators~\cite{Bawaj2015NatCommun}. These approaches often overlook amplification mechanisms, for example those enabled by large Hilbert-space dimensions or high state occupation. Current work in quantum-gravity phenomenology is seeking observables that use these mechanisms to amplify minimal-length effects in scalable dynamical settings.
Quantum theory limits not only observable values but also the speed at which quantum states can evolve~\cite{Deffner2017Review}. Mandelstam and Tamm showed that the minimal time between orthogonal pure states satisfies $\tau_{\mathrm{MT}} \ge \pi\hbar/(2\,\Delta E)$~\cite{Mandelstam1945,QSLFundamentalRobertson1929}. Margolus and Levitin derived the complementary bound $\tau_{\mathrm{ML}} \ge \pi\hbar/\bigl(2\,\langle E\rangle\bigr)$~\cite{Margolus1998}. The QSL is the tightest of these unitary bounds. Geometrically, it is the minimal time to traverse a Bures/Fubini-Study angle in projective Hilbert space and depends only on the first two energy moments, not on full spectral details~\cite{QSLFundamentalBraunsteinCaves1994,QSLFundamentalAnandanAharonov1990,QSLFundamentalJonesKok2010}.  QSLs have been measured in the laboratory. Single-atom Raman-Ramsey interferometry reconstructs the overlap and energy moments from one dataset~\cite{Ness2021SciAdv}, and superconducting-resonator state tomography evaluates the Bures/Fubini--Study angle and energy statistics under software-defined evolution~\cite{Wu2024PRA}. Here we introduce a different, dynamics-centered probe. Prior work focuses on modified propagation or static spectral shifts. We use quantum metrology, specifically quantum speed-limit bounds on state-evolution times, to test how minimal-length/GUP deformations change the geometry of accessible quantum dynamics. This strategy shifts the target from asymptotic travel or polarization properties to intrinsic limits on the pace of transformations in Hilbert space. It provides sensitivity to Planck-suppressed structure without relying on astrophysical emission models or specific dispersion-relation templates.

We incorporate the quadratic GUP deformation into the Mandelstam-Tamm and Margolus-Levitin bounds to obtain testable Planck-scale signatures. We give closed-form, first-order expressions for both bounds in three benchmark systems: (i) a superposition in an infinite square well, (ii) a coherent harmonic-oscillator state, and (iii) a squeezed vacuum state. For any superposition, $\tau_X = \tau_X^{(0)} + \beta\, \tau_X^{(1)}$ with $\tau_X^{(1)} < 0$ for $X=\mathrm{MT}, \mathrm{ML}$. Thus the minimal-length deformation accelerates quantum evolution. We derive a scaling relation, $\delta \tau_X / \tau_X^{(0)} = -\beta\, \mathcal{C}_X\, N_{\mathrm{eff}}^{\,q-k}$. Here $N_{\mathrm{eff}}=\big(\sum_n |c_n|^4\big)^{-1}$, and $q-k>0$ when the GUP-induced matrix elements grow faster than the spectral spacing. We obtain $q-k=1$ for coherent and squeezed states, and $q-k=2$ for equal-weight box superpositions, giving linear or quadratic enhancement with the effective Hilbert-space dimension. Finally, we convert the first-order QSL shifts into bounds on the dimensionless GUP parameter $\beta_{0}$ using demonstrated measurement pipelines. Single-atom Raman-Ramsey data give $\beta_{0,{\rm MT}}^{\max}\!\approx\!6.5\times10^{51}$~\cite{Ness2021SciAdv}. Superconducting-resonator state tomography yields a platform-specific field-mode bound $\beta_{0,{\rm MT}}^{\max}\!\approx\!1.6\times10^{35}$~\cite{Wu2024PRA}. We propose an implementation on the optical-spring mode of a $40\,\mathrm{kg}$ test mass at $f_{\rm eff}\!\approx\!500\,\mathrm{Hz}$. A calibrated coherent displacement $A=1$-$10\,\mathrm{nm}$ and a single-record QSL fit with fractional uncertainty $\varepsilon=10^{-3}$-$\!10^{-4}$ would give $\beta_{0,\mathrm{MT}}^{\max}\!\sim\!10^{3}$-$10^{5}$ from Eq.~(\ref{eq:beta0_QSL_MT}). In our first-order models the ML-based bound is $\simeq 2\times$ tighter. These projections assume two elements: (i) local readout of the detuned, signal-recycled interferometer's optical-spring mode to recover low-frequency sensitivity~\cite{Rehbein2007PRD}, and (ii) back-action-evading or stroboscopic quadrature measurements to keep a near-unitary window and to extract the state overlap and energy moments from a single calibrated record~\cite{Liu2022NJP,Vanner2013NatComm}. The quoted sensitivities are comparable to the best quartz BAW limits~\cite{Campbell2023PRD,Bushev2019PRD} and stronger than sapphire split-bar bounds~\cite{Bourhill2015PRApplied}.

\section{Theoretical Framework}\label{sec:model}
\subsection{Generalized uncertainty principle and minimal-length corrections}
Several quantum-gravity approaches motivate modified uncertainty relations, and thus deformed canonical commutators, including string-theory high-energy scattering \cite{GUPAmati1987,GUPGrossMende1988}, noncommutative or quantized spacetime and deformed Heisenberg algebras \cite{Snyder1947PR,Kempf1995,Doplicher1995CMP,Maggiore1993PLB}, $\kappa$-Poincar\'e and DSR kinematics \cite{AmelinoCamelia2002Nature,AmelinoCamelia2002IJMPD,MagueijoSmolin2002PRL}, and polymer-quantization or loop-inspired dynamics \cite{Ashtekar2003CQG,Hossain2010CQG}. These approaches lead to such modifications; a widely used parametrization is the quadratic GUP
\begin{equation}
[\hat{x},\hat{p}] = i\hbar\bigl(1+\beta\,\hat{p}^{2}\bigr),
\label{eq:GUPcomm}
\end{equation}
Here, the dimensionless parameter \(\beta > 0\) measures the magnitude of quantum-gravitational effects. Physically, this modification sets a fundamental minimum on measurable position uncertainty, typically close to the Planck length.

To simplify our analysis, we use the momentum-space representation introduced by Kempf \textit{et al.}~\cite{Kempf1995}, explicitly implementing the GUP modification as:
\begin{equation}
\hat{p} = \hat{p}_{0}\left(1 + \frac{\beta}{3}\hat{p}_{0}^{2}\right),\quad [\hat{x}_{0}, \hat{p}_{0}] = i\hbar,
\label{eq:momentum_map}
\end{equation}
Here, \(\hat{x}_0\) and \(\hat{p}_0\) are standard canonical operators obeying the usual Heisenberg commutation relation. Assuming the approximation \(\beta  p_{0}^{2}\ll 1\), we expand the Hamiltonian to first order in \(\beta\), producing a universal correction proportional to the fourth power of momentum:
\begin{equation}
\hat{H}_{\beta} = \frac{\hat{p}_{0}^{2}}{2m} + V(\hat{x}_{0}) + \frac{\beta}{3m}\hat{p}_{0}^{4}.
\label{eq:Hbeta}
\end{equation}
Equation~\eqref{eq:Hbeta} recovers known minimal-length corrections for canonical quantum systems frequently studied in quantum-gravity phenomenology, such as the harmonic oscillator~\cite{Bosso2017} and infinite square well~\cite{GUPApplicationQuantumWells2014}. In what follows, we simplify by setting \(\hbar = 1\) and omit higher-order corrections proportional to \(\beta^2\).

\subsection{QSL under minimal-length modifications}

QSL   set a fundamental bound on how rapidly quantum states can evolve. Two key inequalities, derived by  Mandelstam-Tamm (MT)~\cite{Mandelstam1945} and Margolus-Levitin (ML)~\cite{Margolus1998}, offer complementary lower limits on evolution times. The MT bound has the form:
$
\tau_{\mathrm{MT}} \geq {\pi\hbar}/{2\Delta H},
$ 
where $\Delta H$ represents the energy uncertainty. Likewise, the ML bound is:
$
\tau_{\mathrm{ML}} \geq {\pi\hbar}/{2\langle H\rangle },
$ 
where $\langle H\rangle$ is the average energy, and $E_0$ the ground-state energy. Recently, these two bounds have been combined into a unified expression~\cite{Deffner2017Review,Maleki2023SpeedLimitMetrology}:
\begin{equation}\label{eq:UnifiedQSL}
\tau_{\mathrm{QSL}}(\mathcal{L}) = \max\left\{\frac{\hbar\,\mathcal{L}}{\Delta H}, \frac{2\hbar\,\mathcal{L}^{2}}{\pi\langle H\rangle}\right\},
\end{equation}
Here, the Bures angle $\mathcal{L}(t) = \arccos|\braket{\psi_{0}}{\psi(t)}|$ measures how distinguishable two quantum states are.

To study how minimal-length modifications affect QSL, we apply Eq.~\eqref{eq:UnifiedQSL} to three illustrative classes of quantum states, selected for theoretical simplicity and experimental feasibility: (i) superpositions of eigenstates in an infinite square well, (ii) coherent harmonic oscillator states, and (iii) squeezed vacuum states. Each of these states evolves under the minimal-length Hamiltonian given by Eq.~\eqref{eq:Hbeta}. Expanding the QSL up to first order in the small parameter $\beta$, we have:
\begin{equation}
\tau_{\mathrm{QSL}} = \tau_{\mathrm{QSL}}^{(0)}\left[1+\beta\,\tau_{1}+\mathcal{O}(\beta^{2})\right],
\end{equation}
where $\tau_{\mathrm{QSL}}^{(0)}$ is the standard (undeformed) QSL, and $\tau_{1}$ represents the first-order minimal-length correction. The correction term $\tau_{1}$ explicitly depends on the details of the quantum state and the geometry of the system.

\section{QSL Corrections: Particle in an Infinite Square Well}\label{sec:box}
We begin by studying minimal-length corrections to QSL using the infinite square-well model, selected for its simplicity and broad relevance in quantum physics. Using natural units ($\hbar = m = 1$), the unperturbed Hamiltonian is given by
\begin{equation}
\hat{H}_0 = \frac{\hat{p}^2}{2} + V(x), \quad V(x) =
\begin{cases}
0, & 0 < x < L,\\[2pt]
\infty, & \text{otherwise},
\end{cases}
\end{equation}
with eigenstates and eigenenergies 
\begin{equation}\label{eq:pb}
\psi_n^{(0)}(x) = \sqrt{\frac{2}{L}}\sin\left(\frac{n\pi x}{L}\right),\quad
E_n^{(0)} = \frac{n^2\pi^2\hbar^2}{2L^2}.
\end{equation}

To incorporate minimal-length effects, we supplement Eq. (\ref{eq:pb}) with the leading GUP contribution; namely, the quartic-momentum term that appears in Eq. (\ref{eq:Hbeta}). Applying perturbation theory yields the corrected energy levels:
\begin{equation}
E_n = E_n^{(0)} + E_n^{(1)} = \frac{n^2 \pi^2 }{2 L^2} + \frac{1}{4}\beta\left(\frac{n\pi}{L}\right)^4.
\end{equation}
This clearly demonstrates that minimal-length effects shift energy levels upward, especially at higher energies.

We now analyze QSL corrections for a general superposition of energy eigenstates:
\begin{equation}
\ket{\Psi(t)} = \sum_{n} c_n \, e^{-iE_n t/\hbar}\ket{\psi_n^{(0)}}, \quad \sum_n |c_n|^2 = 1.
\end{equation}

In this setting, we analytically derive first-order corrections to the  MT and ML bounds. These corrections, derived in detail in Appendix~\ref{PIB}, are given by:
\begin{align}\label{BOXMT}
\tau_{\mathrm{MT}}^{(1)} = -\frac{\hbar}{\Delta E^{(0)}}\,
\frac{\Re\!\bigl[F^{(0)*}(t)F^{(1)}(t)\bigr]}
     {\sqrt{1-|F^{(0)}(t)|^2}\,|F^{(0)}(t)|}\nonumber\\ 
-\frac{\hbar\,\tau_{\mathrm{MT}}^{(0)}}{(\Delta E^{(0)})^2}\,\Delta E^{(1)},
\end{align}
 and
\begin{equation}\label{BOXML}
\tau_{\mathrm{ML}}^{(1)} = 
\frac{\pi\hbar}{\bar{E}^{(0)}}\mathcal{L}^{(0)}(t)\mathcal{L}^{(1)}(t)
-\frac{\pi\hbar}{2(\bar{E}^{(0)})^2}\bigl(\mathcal{L}^{(0)}(t)\bigr)^2\bar{E}^{(1)}.
\end{equation}

\begin{figure}[htbp]
  \centering
  \includegraphics[width=\linewidth]{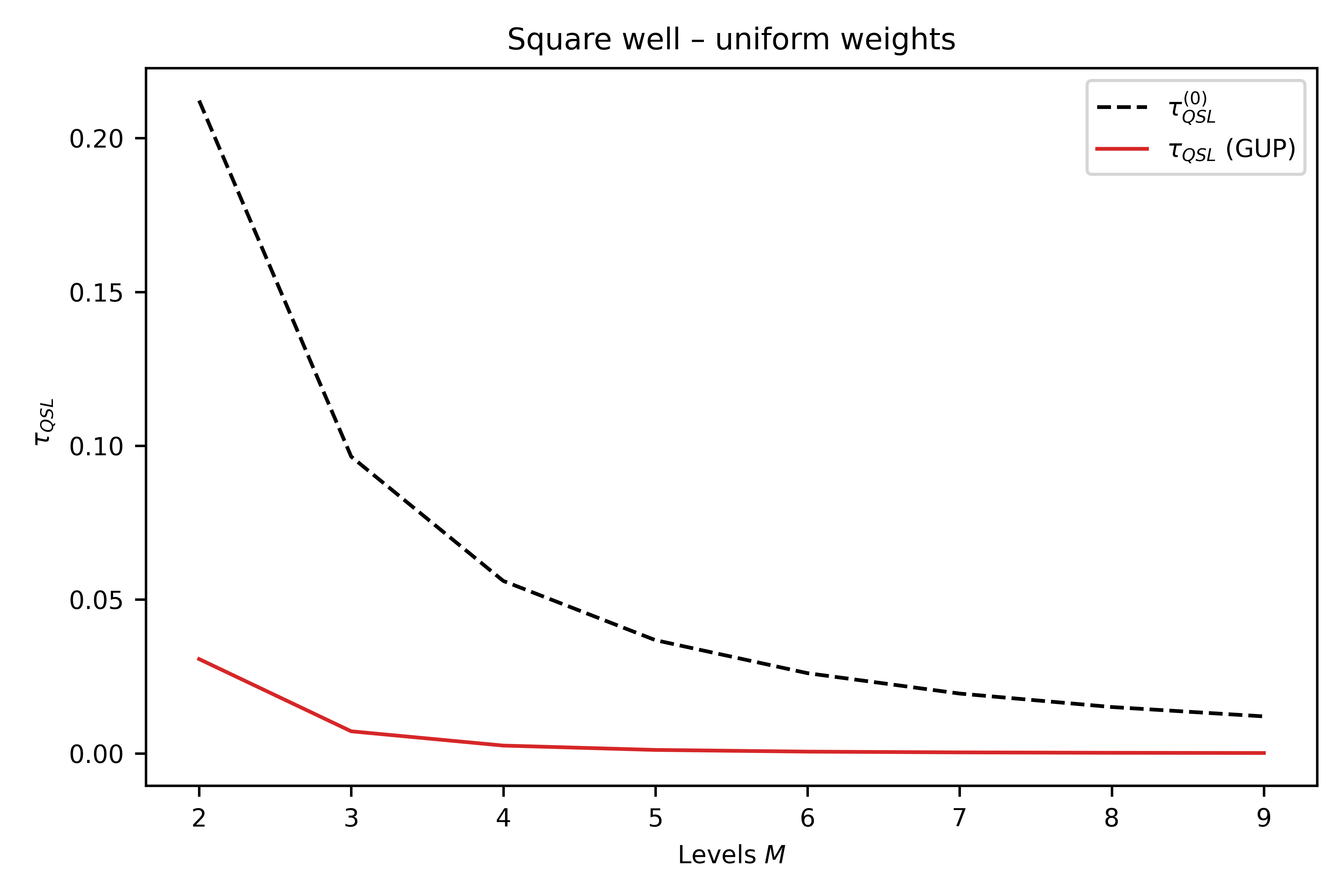}%
\caption{ QSL $\tau_{\min}$ for a particle in an infinite square well, with equally occupied energy levels. The dashed curve shows the undeformed limit $\tau_{\min}^{(0)}$, while the solid curve includes corrections from the quadratic GUP with $\beta=1.0\times10^{-2}$.}

  \label{fig:QSLvsM}
\end{figure}

\begin{figure}[htbp]
  \centering
  \includegraphics[width=\linewidth]{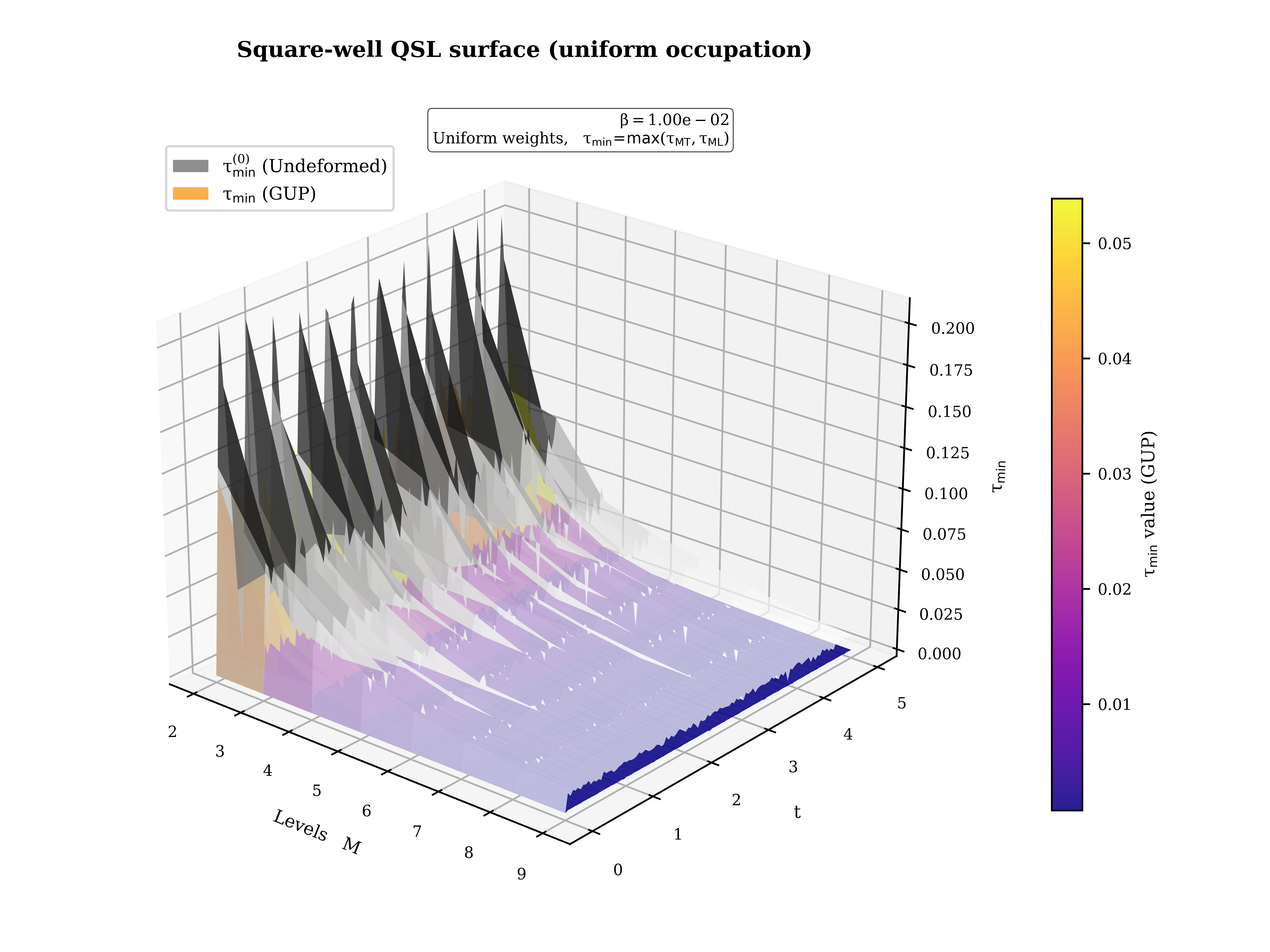}%
\caption{ QSL $\tau_{\min}(M,t)$ as a function of both the number of equally occupied energy levels $M$ and evolution time $t$. The gray mesh represents the undeformed result $\tau_{\min}^{(0)}(M,t)$, while the colored mesh (viridis colormap) shows the GUP-corrected values at $\beta=1.0\times10^{-2}$. The visible separation between these surfaces clearly illustrates the acceleration of quantum dynamics due to minimal-length effects.}
\label{fig:QSLsurface}
\end{figure}

Figure~\ref{fig:QSLvsM} shows the QSL as a function of the number of equally occupied energy levels $M$ (effective dimension $N_{\mathrm{eff}}=M$). The dashed curve represents the undeformed result $\tau_{\min}^{(0)}\propto M^{-1}$, while the solid curve shows the value corrected by the quadratic GUP. For every $M$, the corrected bound is consistently below the undeformed bound, supporting the universal acceleration described in Sec.~\ref{sec:box}. The increasing gap clearly illustrates the predicted linear amplification, $\delta\tau_{\min}\propto\beta N_{\mathrm{eff}}$.

Figure~\ref{fig:QSLsurface} shows how these results change over evolution time $t$. The gray surface represents the undeformed limit $\tau_{\min}^{(0)}(M,t)$, while the colored surface shows the GUP-corrected limit $\tau_{\min}(M,t)$.  These numerical results support our analytical predictions. Minimal-length effects accelerate quantum evolution, and importantly, the effective Hilbert-space dimension significantly amplifies this acceleration. We present detailed analytical derivations in Appendix~\ref{PIB}.

\section{QSL for the GUP-Deformed Harmonic Oscillator}\label{sec:sho_coherent}

We now explore how minimal-length corrections affect QSL for a harmonic oscillator. We focus on a one-dimensional harmonic oscillator modified by the GUP, introducing a quartic momentum correction. The modified Hamiltonian is given by:
\begin{equation}
\hat{H} = \hat{H}_0 + \frac{\beta}{3}\hat{p}^4, \quad \text{with} \quad \hat{H}_0 = \frac{\hat{p}^2}{2} + \frac{\hat{x}^2}{2},
\end{equation}
Here, $\beta$ is a small, dimensionless parameter that measures the strength of the minimal-length correction from GUP. Applying first-order Rayleigh-Schr\"odinger perturbation theory, we obtain corrections to the energy eigenvalues as~\cite{FormalismPlanckHO2017}:
\begin{equation}\label{eq:sho_energy_levels}
E_n = E_n^{(0)} + \frac{\beta}{12}(6n^2 + 6n + 3),\quad E_n^{(0)} = n + \frac{1}{2}.
\end{equation}
The unperturbed eigenstates $\ket{n_0}$ are the standard harmonic oscillator number states. To first order in $\beta$, we express the perturbed eigenstates as:
\begin{equation}\label{eq:sho_perturbed_states}
\ket{n} = \ket{n_0} - \frac{\beta}{12}\ket{n_1} + \mathcal{O}(\beta^2),
\end{equation} Here, $\ket{n_1}$ involves a sum over intermediate states of opposite parity; the explicit derivation appears in Appendix~\ref{sec:app_sho}.
In terms of the number operator $N$, the Hamiltonian simplifies to:
\begin{equation}\label{eq:sho_H_number_main}
\hat{H} = N + \frac{1}{2} + \frac{\beta}{12}(6N^2 + 6N + 3).
\end{equation}

\subsection{QSL Corrections for Coherent States}

We now consider coherent states, widely used in quantum metrology and optics for their classical-like behavior and minimal uncertainty. We consider a coherent state $\ket{\alpha}$ that evolves according to the GUP-modified harmonic oscillator Hamiltonian. The time-evolved coherent state is given by:
\begin{equation}\label{eq:sho_coherent_time_state_main}
\ket{\alpha'(t)} = e^{-\frac{|\alpha_0|^2}{2}}\sum_{n=0}^{\infty}\frac{\alpha_0^n}{\sqrt{n!}}\,e^{-iE_n t}\ket{n},
\end{equation}
Here, the perturbed eigenenergies $E_n$ follow from Eq.~\eqref{eq:sho_energy_levels}. Using this expression, we derive analytical formulas for the expectation values and fidelity amplitude; full details are provided in the Appendix~\ref{sec:app_sho}. To first order in $\beta$, the results are:
\begin{subequations}\label{eq:sho_expectation_coherent_main}
\begin{align}
\langle H\rangle_{\alpha'} &= \frac{1}{2} + \alpha_0^2 + \frac{\beta}{4}\left(1+4\alpha_0^2+2\alpha_0^4\right),\\[3pt]
(\Delta H_{\alpha'})^2 &= \alpha_0^2 + 2\beta\left(\alpha_0^2+\alpha_0^4\right),\\[3pt]
|\braket{\alpha'(0)}{\alpha'(t)}| &= e^{\alpha_0^2(\cos t - 1)}\nonumber\\
&\quad\times\left[1-\beta\,\alpha_0^2 t\,(1+\alpha_0^2\cos t)\sin t\right].
\end{align}
\end{subequations}
From these analytical results, we derive concise first-order corrections to the  MT and  ML bounds:
\begin{align}\label{eq:sho_MT_coh_main}
\tau_{\text{MT}} &= \frac{S_0(t)}{\alpha_0}-\beta\left[\frac{(1+\alpha_0^2)}{\alpha_0}S_0(t)-\alpha_0D(t)\right],\\[5pt]
\tau_{\text{ML}} &= \frac{2\,S_0^2(t)}{\pi\left(\frac{1}{2}+\alpha_0^2\right)}-\beta\,\frac{(1+4\alpha_0^2+2\alpha_0^4)S_0^2(t)}{\pi(1+2\alpha_0^2)^2}\nonumber\\[3pt]
&\quad + \frac{4\beta\alpha_0^2(1+2\alpha_0^2)S_0(t)D(t)}{\pi(1+2\alpha_0^2)^2}.
\label{eq:sho_ML_coh_main}
\end{align}

We introduce auxiliary functions for clarity:
\begin{align}
S_0(t) &= \arccos\left[e^{\alpha_0^2(\cos t - 1)}\right],\\[3pt]
D(t) &= \frac{t(1+\alpha_0^2\cos t)\sin t\, e^{\alpha_0^2(\cos t - 1)}}{\sqrt{1-e^{2\alpha_0^2(\cos t - 1)}}}.
\end{align}

\begin{figure}[htbp]
  \centering
  \includegraphics[width=\linewidth]{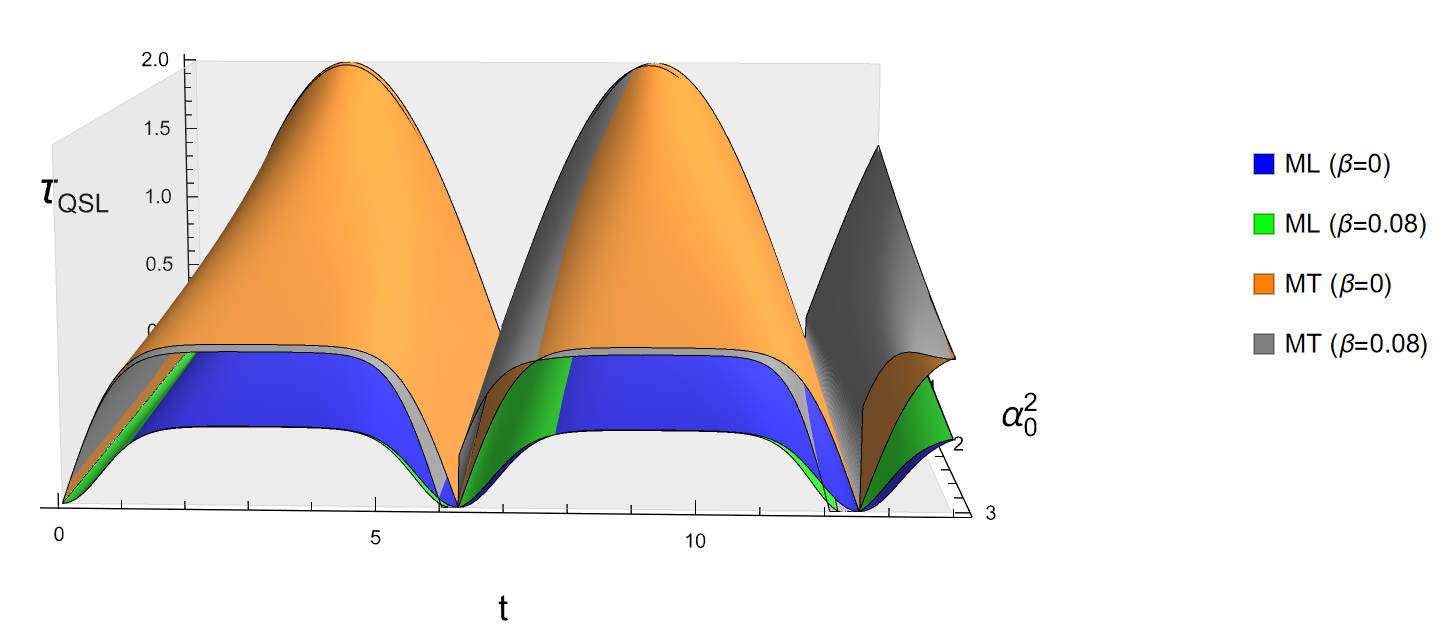}
\caption{\textbf{Impact of quantum gravity on MT and ML bounds for coherent states.}
The plot shows QSL surfaces versus evolution time \( t \) and mean photon number \( \alpha_0^2 \). The orange and blue surfaces show the MT and ML bounds without GUP corrections (\( \beta = 0 \)). The gray and green surfaces include GUP corrections (\( \beta = 0.08 \)). GUP corrections shorten the minimal evolution time, clearly accelerating quantum evolution.}
\label{fig:plot_beta_coh}
\end{figure}

\begin{figure}[htbp]
  \centering
  \includegraphics[width=\linewidth]{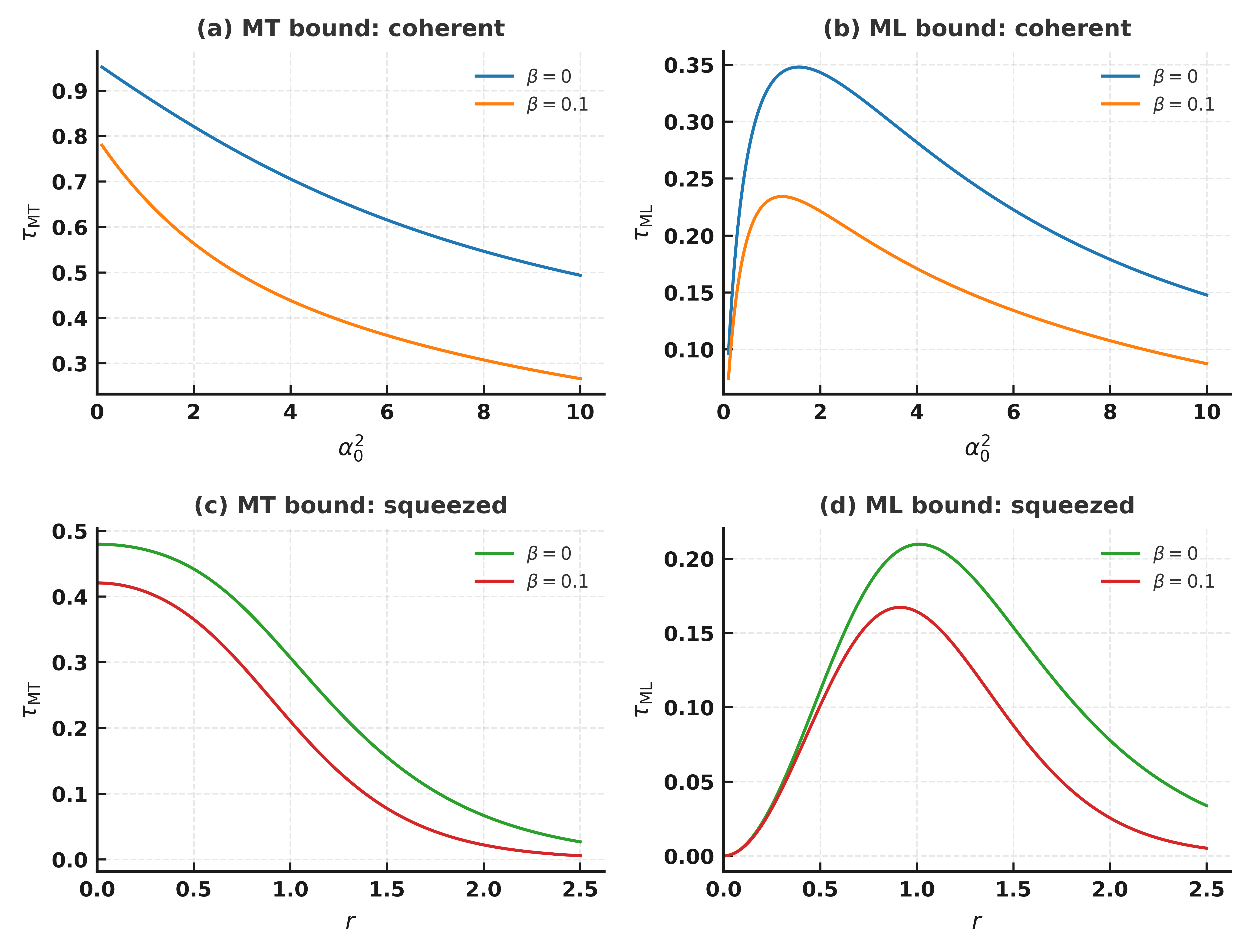}
 \caption{\textbf{Minimal-length acceleration of quantum evolution.}
This figure presents  MT and ML bounds for a fixed Bures angle \( \mathcal{L} = 1 \). Panels (a) and (b) display the MT and ML bounds for GUP-modified coherent states, plotted against mean photon number \( n = \alpha_{0}^{2} \). Panels (c) and (d) show the same bounds for GUP-modified squeezed vacuum states versus the squeezing parameter \( r \). In each panel, black curves show results without deformation (\( \beta = 0 \)), and colored curves include the quartic \( p^{4} \) corrections at \( \beta = 0.1 \). For both state types, minimal-length effects shorten the GUP time.}
 \label{fig:QSL_fourpanel}
\end{figure}

Figures~\ref{fig:plot_beta_coh} and~\ref{fig:QSL_fourpanel} show that these GUP corrections consistently lower both MT and ML bounds, accelerating quantum dynamics. This acceleration strengthens as the coherent amplitude $\alpha_0$ increases, demonstrating a clear amplification mechanism. Physically, the minimal-length effect widens the momentum distribution. This widening increases the energy variance, shortening the minimal evolution time as described by the MT bound. These results have important practical implications for quantum-optical and optomechanical experiments. Since coherent states are easy to create and control experimentally, our results highlight coherent-state-based QSL measurements as promising tools for probing minimal-length quantum gravity effects.

\subsection{GUP corrections to QSL: Squeezed states}\label{sec:squeezed_states}

Next, we study how minimal-length modifications from the GUP influence QSL in squeezed vacuum states. Squeezed states play a key role in quantum metrology, notably in gravitational-wave detection and quantum sensors~\cite{Abbott2016,Schnabel2017}, because they enhance sensitivity to quantum fluctuations. Under first-order GUP corrections, the squeezed vacuum state evolves in time as~\cite{FormalismPlanckHO2017}:
\begin{equation}\label{eq:sq_main}
\ket{\varsigma(t)}=\sum_{n=0}^{\infty}f(2n)e^{-iE_{2n}t}\ket{2n},
\end{equation}
where the amplitudes
\begin{equation}
f(2n)=\frac{\sqrt{(2n)!}\,(-\tanh r)^n}{2^n n!\sqrt{\cosh r}}
\end{equation}
remain unchanged at this approximation level. However, first-order GUP corrections modify the eigen-energies, as explicitly derived in Appendix~\eqref{sec:app_sho}. We calculate the Hamiltonian's expectation value and variance, including first-order GUP corrections, obtaining:
\begin{subequations}
\label{eq:sq_energy_main}
\begin{align}
\langle H\rangle_{\varsigma}&=\frac{1}{2}\cosh(2r)+\frac{\beta}{16}[1+3\cosh(4r)],\\[4pt]
\Delta H_{\varsigma}^{2}&=2\,\cosh^2 r\,\sinh^2 r+\frac{3\beta}{4}\sinh(2r)\sinh(4r).
\end{align}
\end{subequations}

Additionally, we explicitly derive the fidelity amplitude-the overlap between initial and evolved squeezed states-in Appendix~\ref{sec:app_sho}, giving:
\begin{equation}\label{eq:app_fidelity_expanded1}
|\braket{\varsigma(0)}{\varsigma(t)}|= \frac{\sqrt{2}\,\sech r}{D^{1/4}(t)} + \beta\frac{8t\,\cosh^5 r\,\sinh^2 r\,G(t)}{D^2(t)\,[1-2\cos t\,\tanh^2 r+\tanh^4 r]^{1/4}},
\end{equation}
with auxiliary functions:
\begin{subequations}
\begin{align}
D(t)&=3+\cosh(4r)-2\cos t\,\sinh^2(2r),\\[4pt]
G(t)&=-4\sin t+\sin(2t)\tanh^2 r+2\sin t\,\tanh^4 r.
\end{align}
\end{subequations}

The angular distance (Bures angle) between the initial and evolved states is defined as:
\begin{equation}\label{eq:bures_squeezed_main}
S(t)=\arccos|\braket{\varsigma(0)}{\varsigma(t)}|.
\end{equation}

From Eqs.~\eqref{eq:sq_energy_main} and \eqref{eq:bures_squeezed_main} the MT bound follows directly (closed form given in Appendix~\ref{sec:app_sho}). 
\begin{equation}\label{eq:MT_squeezed_main}
\tau_{\mathrm{MT}}
=\frac{\hbar S_0(t)}{\Delta H_0}
+\beta\hbar\left[
\frac{S_1(t)}{\Delta H_0}
-\frac{S_0(t)\Delta H_1}{(\Delta H_0)^{2}}
\right]
.
\end{equation}
Similarly, the ML bound acquires an explicit first-order correction:
\begin{equation}\label{eq:ML_squeezed_main}
\begin{aligned}
\tau_{\text{ML}}&=\frac{4S^2(t)}{\pi\cosh(2r)}+\frac{\beta\,S(t)}{2\pi\cosh(2r)}\\[4pt]
&\quad\times\left[K(r)\,S(t)+\frac{32t\cosh^5 r\sinh^2 r\,H(t)}{R^{7/4}(t)P^{1/4}(t)Z(t)}\right],
\end{aligned}
\end{equation}
with
\begin{equation}
K(r)=(1+3\cosh(4r))\sech(2r).
\end{equation}
Numerical evaluation clearly shows that minimal-length corrections consistently lower both MT and ML bounds (Fig.~\ref{SQP}). The periodic structure of the QSL curves persists, but GUP corrections notably shorten evolution times (Fig.~\ref{fig:QSL_fourpanel}, panels c-d). Physically, this acceleration suggests that minimal-length effects could relax fundamental timing constraints in high-precision interferometric setups that employ squeezed states, such as gravitational-wave detectors. Practically, this acceleration enhances temporal resolution and potentially improves quantum-enhanced measurement sensitivity. However, this benefit comes with stricter demands on decoherence management and precise timing control. Thus, squeezed states stand out as ideal candidates for probing Planck-scale physics in realistic laboratory setups.
Our results underscore GUP measurements using squeezed states as powerful tools to detect quantum-gravity effects, bridging fundamental theory and current quantum-optical experiments.

\begin{figure}[htbp]
  \centering
  \includegraphics[width=\linewidth]{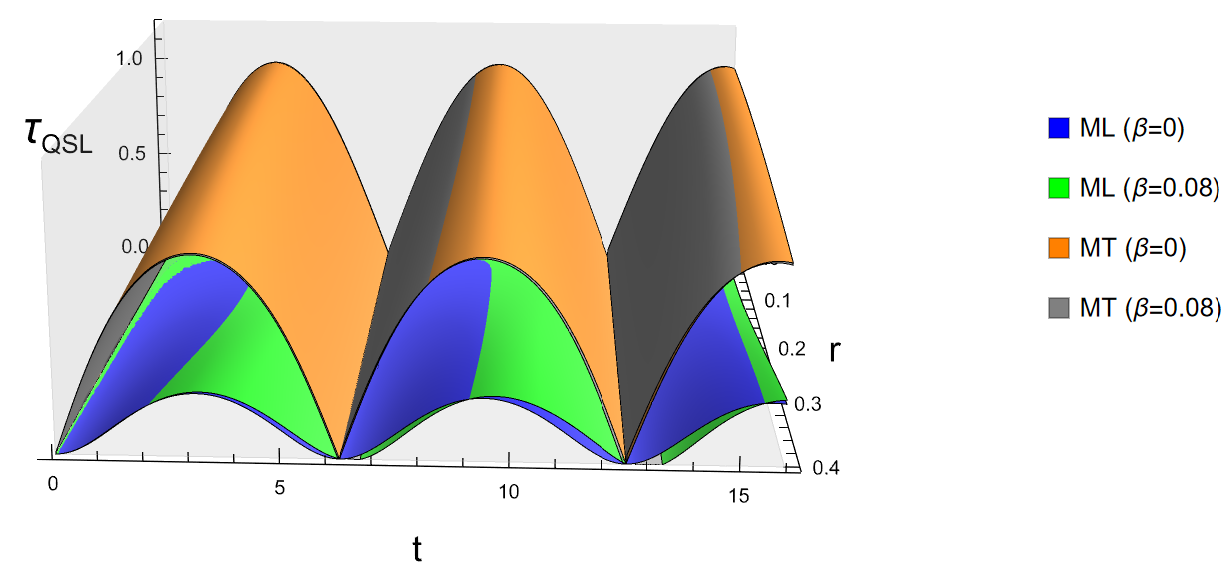}
\caption{\textbf{Acceleration of quantum dynamics in squeezed vacuum states from minimal-length effects.} 
This figure plots the  MT and ML bounds against the squeezing parameter $r$ and evolution time $t$. As shown, the GUP-corrected bounds (gray for MT, green for ML; $\beta=0.08$) fall below the corresponding undeformed surfaces (orange for MT, blue for ML).}
\end{figure}

\section{First-Order Universal Scaling of QSL}
\label{SQP}
We next demonstrate, and prove analytically in Appendix~\ref{A} and Appendix~\ref {B}, that the first-order GUP correction to the  MT and ML bounds follows a universal scaling with the effective Hilbert-space size.  For an arbitrary pure state, the fractional shift is
\begin{equation}
  \frac{\delta\tau_X}{\tau_X^{(0)}}=-\beta\,\mathcal F_X(N_{\mathrm{eff}}),
  \qquad X\in\{\mathrm{MT},\mathrm{ML}\},
\end{equation}
where $\beta$ is the deformation parameter, $N_{\mathrm{eff}}=(\sum_n|c_n|^{4})^{-1}$, and explicit forms of $\mathcal F_X$ appear in Eqs.~(A10)-(A11).  For a coherent state, $N_{\mathrm{eff}}=\bar n=|\alpha|^{2}$, the shifts read 
\begin{align}
\left|\frac{\delta\tau_{\mathrm{MT}}}{\tau_{\mathrm{MT}}^{(0)}}\right| &=2(\bar n+1)\beta,\nonumber\\
\left|\frac{\delta\tau_{\mathrm{ML}}}{\tau_{\mathrm{ML}}^{(0)}}\right| &=\frac{\bar n^{2}+2\bar n+\tfrac12}{\bar n+\tfrac12}\beta\label{HO2},
\end{align}\label{HO1}
which simplify to $2\bar n\beta$ and $\bar n\beta$ when $\bar n\gg1$.  For a squeezed vacuum with $N_{\mathrm{eff}}=\sinh^{2}r$ one obtains 
\begin{align}
\left|\frac{\delta\tau_{\mathrm{MT}}}{\tau_{\mathrm{MT}}^{(0)}}\right|&=\frac{4\bar n^{2}+9\bar n+4}{\bar n+1}\beta,\nonumber\\
\left|\frac{\delta\tau_{\mathrm{ML}}}{\tau_{\mathrm{ML}}^{(0)}}\right|&=\frac{2\bar n^{2}+2\bar n+\tfrac12}{\bar n+\tfrac12}\beta,
\end{align}\label{HOS}
so that for $\bar n\gg1$ the MT and ML corrections scale as $6\bar n\beta$ and $3\bar n\beta$, respectively.  Finally, for an equal-weight superposition of the first $M$ infinite-square-well eigenstates, where $N_{\mathrm{eff}}=M$, one finds the quadratic enhancement discussed in Appendix \ref{A}
For a uniform superposition of the first $M$ eigenstates of the infinite square well ($N_{\mathrm{eff}} = M$), the first-order GUP shifts are  
\begin{align}
\bigl|\delta\tau_{\mathrm{MT}}/\tau_{\mathrm{MT}}^{(0)}\bigr|
   &= \frac{\beta\,\varepsilon\,(2M+1)}{7(M-1)(16M^{2}+30M+11)} \times \nonumber\\
   &\quad(48M^{4}+75M^{3}-70M^{2}-90M+37), \\
\bigl|\delta\tau_{\mathrm{ML}}/\tau_{\mathrm{ML}}^{(0)}\bigr|
   &= \beta\,\varepsilon\,\frac{(M+1)(2M+1)(3M^{2}+3M-1)}{5(2M^{2}+3M-5)}.
\end{align}

In the large-$M$ limit these expressions simplify to
$\beta\,\varepsilon\,(6M^{2}/7)$ for the MT bound and $\beta\,\varepsilon\,(3M^{2}/5)$ for the ML bound.
\begin{figure}[tbp]          
  \centering
  \includegraphics[width=\columnwidth]{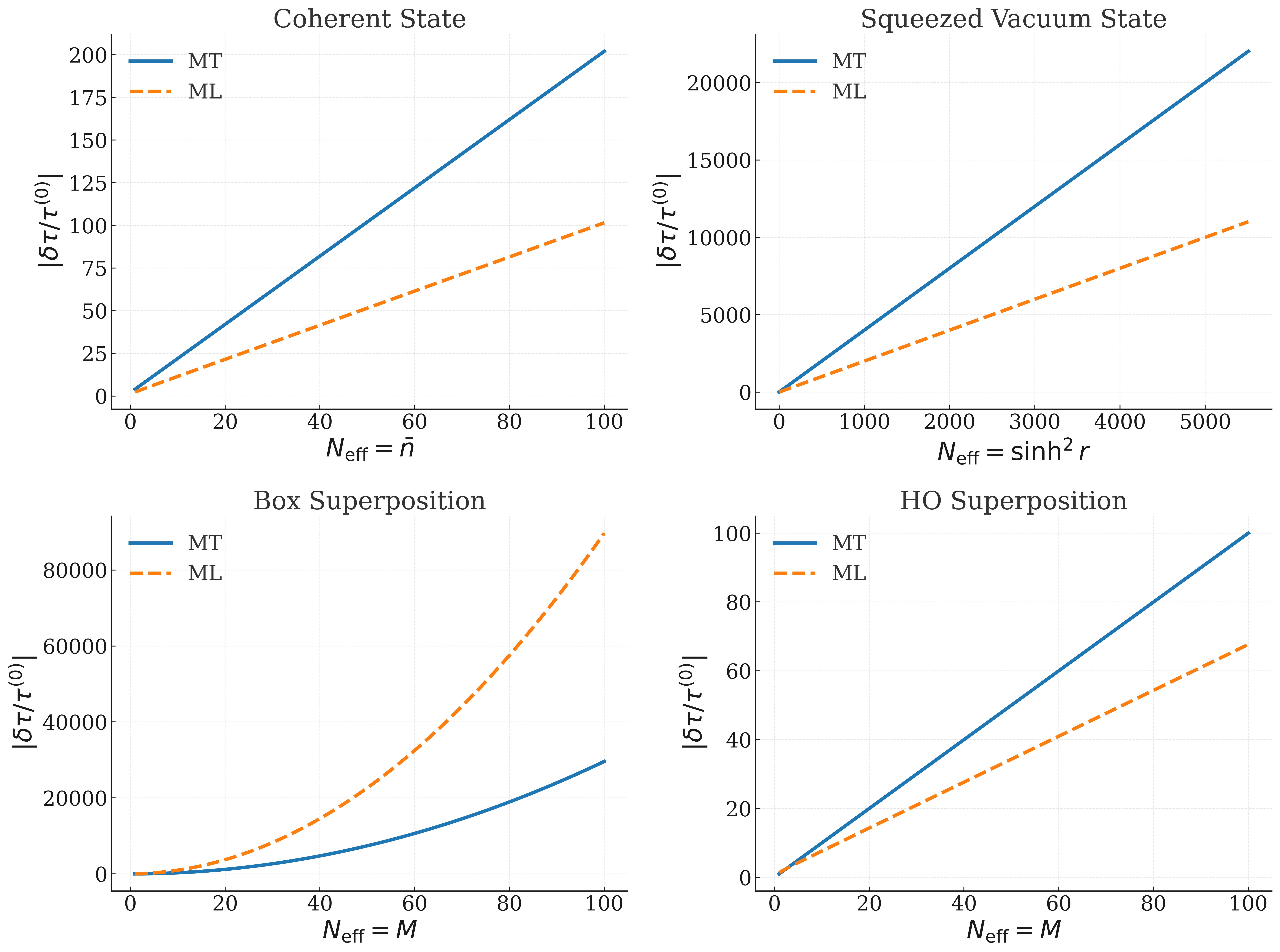}
  \caption{%
    First-order fractional quantum-speed-limit shifts
    $\bigl|\delta\tau/\tau^{(0)}\bigr|$ for Mandelstam-Tamm (solid blue)
    and Margolus-Levitin (dashed orange) bounds as functions of the
    effective Hilbert-space size $N_{\mathrm{eff}}$.  Panels:
    (a) coherent state ($N_{\mathrm{eff}}=\bar n$);
    (b) squeezed vacuum ($N_{\mathrm{eff}}=\sinh^{2} r$);
    (c) box superposition ($N_{\mathrm{eff}}=M$); and
    (d) equal-weight harmonic-oscillator superposition
    ($N_{\mathrm{eff}}=M$).  All curves are plotted for $\beta=1$ to
    highlight the universal algebraic scaling.}
  \label{fig:QSL-scaling}
\end{figure}
Hence, box superpositions exhibit a quadratic enhancement with effective dimension (scaling as $\beta\,\varepsilon\,M^{2}$), whereas coherent and squeezed oscillator states grow only linearly, underscoring the ability of QSL observables to magnify minimal-length effects beyond static spectroscopic shifts. Figure~\ref{fig:QSL-scaling} compares these analytic trends with full numerics over the
entire $N_{\mathrm{eff}}$ range and finds quantitative agreement.

\section{Asymptotic power law for generic ladders}\label{ssec:generic-scaling}
The linear \(N_{\mathrm{eff}}\) growth seen for coherent, squeezed, and square-well states is just one instance of a broader large-\(N_{\mathrm{eff}}\) scaling law, derived in Appendix \ref{B} (Eq. \eqref{BE}).  
Consider a ladder with \(E^{(0)}_{n}\!\propto n^{k}\) and diagonal GUP elements \(W_{n}\!\propto n^{q}\).  
For a uniform superposition of the first \(N_{\mathrm{eff}}\) levels we obtain  
\begin{equation}
\frac{\delta\tau_{X}}{\tau_{X}^{(0)}}=-\beta\,\mathcal{C}_{X}(k,q)\,N_{\mathrm{eff}}^{\,q-k},
\qquad
X\in\{\text{MT},\text{ML}\},
\label{eq:generic-scaling-main}
\end{equation}
where the coefficients \(\mathcal{C}_{\mathrm{ML}}\) and \(\mathcal{C}_{\mathrm{MT}}\) are given in Eq. \eqref{BE1}.  
When \(q>k\), the GUP term grows faster with \(n\) than the unperturbed spectrum, so the factor \(N_{\mathrm{eff}}^{\,q-k}\) amplifies the shift algebraically.  
The size of this enhancement is set by the number of occupied levels.
Equation~\eqref{eq:generic-scaling-main} matches our three test cases.  
For the harmonic oscillator, \((k,q)=(1,2)\) yields the linear dependence \(\propto\beta\bar n\).  
For the infinite square well, \((k,q)=(2,4)\) produces a quadratic growth \(\propto\beta N_{\mathrm{eff}}^{2}\). A deep Morse ladder, \((k,q)=(2,3)\), lies between these limits and scales linearly with \(\beta N_{\mathrm{eff}}\). If \(q\le k\), the first-order term no longer grows with system size, so the speed limits lose sensitivity to~\(\beta\).  Off-diagonal GUP terms appear only at second order; they leave the scaling exponent unchanged. Overall, Eq.~\eqref{BE} provides a practical rule of thumb for estimating the Planck-scale reach of large, finite-bandwidth quantum systems.

\section{Experimental quantum speed limits and bounds on the GUP parameter}
\label{sec:ExpQSLtoGUP}

{Quantum speed limits (QSLs) have been measured in the laboratory on single atoms and superconducting resonators \cite{Ness2021SciAdv,Wu2024PRA}. In these experiments the QSL is obtained from the state overlap and the energy moments reconstructed from the \emph{same} measurement record. No auxiliary clock is required \cite{Deffner2017Review}. We use this to constrain the quadratic generalized-uncertainty-principle (GUP) parameter \( \beta_0 \) by mapping measured short-time deviations from the closed-system Mandelstam-Tamm (MT) baseline to first order (derivation in App.~A). This section reviews QSL evaluation in practice, states the GUP\(\to\)QSL mapping, applies it to two datasets, and gives conservative projections for macroscopic oscillators. The resulting bounds are compiled in Table~\ref{tab:beta0_bounds_outlook}.

In single-atom Raman-Ramsey interferometry, a short Raman pulse prepares two internal-state branches. One branch is displaced in a near-harmonic trap, and a Ramsey fringe is reconstructed after a programmable delay. Fitting the phase and visibility yields the complex overlap and the energy moments from the same record \cite{Ness2021SciAdv}. In superconducting microwave resonators, tailored photonic states evolve freely for programmed delays. Full quantum-state tomography reconstructs \(\rho_t\), the photon statistics, and therefore \(\mathcal F(t)\) and the energy moments \cite{Wu2024PRA}. In both platforms the observables \(\{\mathcal F(t),\langle H\rangle,\Delta E\}\) are internal to a single system, which suppresses clock-synchronization loopholes.

For coherent mechanical states with \(x_{\rm zpf}=\sqrt{\hbar/(2m\omega)}\) and \(\omega=2\pi f\), the mean phonon number and the energy moments are
\begin{equation}
\label{eq:coherent_moments}
\begin{gathered}
\bar n = \frac{A^2}{4x_{\rm zpf}^2} = \frac{m\omega A^2}{2\hbar},\\
\Delta E = \hbar\omega\sqrt{\bar n},\qquad
\langle H\rangle - E_0 = \hbar\omega\,\bar n .
\end{gathered}
\end{equation}

We fit the short-time Bures angle $\mathcal{L}(t)=\arccos\sqrt{\mathcal{F}(t)}$ to the Mandelstam-Tamm (MT) baseline $\mathcal{L}^{(0)}(t)=(\Delta E/\hbar)\,t$ to extract the QSL time $\tau$, and summarize its precision by a single fractional uncertainty $\varepsilon \equiv |\delta\tau|/\tau^{(0)}$. For coherent states, $\varepsilon$ is dominated by the calibration of $(A,m,\omega)$ and the sampling-time jitter $\delta t_{\mathrm{rms}}$. Explicit uncertainty-propagation formulas (and the Margolus-Levitin exponents) are given in App. \eqref{A}.
For a single mechanical mode with mass \(m\), angular frequency \(\omega\), and coherent peak displacement \(A\), the quadratic GUP kinetic deformation \( \Delta \hat H \propto \beta_0\,\hat p^4/m \) gives the first-order coherent-state shift
\begin{equation}
\Big(\tfrac{\delta\tau}{\tau}\Big)_{\!\rm MT}\simeq
\beta_0\,\frac{4m^2\omega^2A^2}{(M_P c)^2},
\label{eq:shift_MT}
\end{equation}
which leads to the bound
\begin{equation}
\boxed{\;
\beta_{0,\mathrm{MT}}^{\max}=\frac{\varepsilon\,(M_P c)^2}{4\,m^{2}\omega^{2}A^{2}}\;}
\label{eq:beta0_QSL_MT}
\end{equation}

upon identifying the experimental allowance with \(\varepsilon \equiv |\delta\tau_0|/\tau^{(0)}\). For coherent states the Margolus-Levitin mapping differs by a factor of two. Hence \(\beta_{0,{\rm ML}}^{\max}=2\,\beta_{0,{\rm MT}}^{\max}\).\footnote{For coherent states, the ML mapping differs from MT by an exact factor of two at first order in the quadratic deformation derived here.} The normalization is \((M_P c)^2=4.2573\times10^{1}\ \mathrm{kg^2\,m^2\,s^{-2}}\), so \(\beta_0\) is dimensionless.

For a single LC field mode with impedance \(Z=\sqrt{L/C}\) we identified \(P\to Q\) and \(m\to C\). Using \(\Phi_{\rm zpf}=\sqrt{\hbar Z/2}\) and \(Q_{\rm zpf}=\sqrt{\hbar/(2Z)}\), the dimensionless perturbation that multiplies \(W=\hat p^4/4\) is \(\beta_{\rm diml}=\beta_0\hbar/[4Z(M_P c)^2]\). A momentum-displaced coherent state then yields
\begin{equation}
\boxed{\;
\beta_{0,{\rm MT}}^{\max}\simeq \frac{2Z\,(M_P c)^2}{\hbar}\,\frac{\varepsilon}{\bar n}\;}
\label{eq:beta0_LC}
\end{equation}

Superconducting-resonator QSL data illustrate the substitution.
The experiment reconstructed a single microwave mode at \(f_{\rm eff}=4.0~\mathrm{MHz}\) with coherent amplitudes \(\alpha=0.4,1.0,2.0\), giving \(\bar n=0.16,1,4\)~\cite{Wu2024PRA}.
Using \(\tau_{\rm MT}^{(0)}=\pi/(2\omega_{\rm eff}\sqrt{\bar n})\) with \(\omega_{\rm eff}=2\pi f_{\rm eff}\) and timing steps \(\delta t=0.5\)--\(5\)~ns, we obtain \(\varepsilon=0.016\) for \(Z=50~\Omega\), \(\bar n=4\), and \(\delta t=0.5\)~ns.
Equation~\eqref{eq:beta0_LC} gives \(\beta_{0,{\rm MT}}^{\max}\approx 1.6\times 10^{35}\).
The scaling is linear in \(Z\) and \(\delta t\), and inverse in \(\bar n\).

Single-atom Raman-Ramsey data provide a second benchmark.
The experiment observed MT/ML crossover with period \(T\simeq 16~\mu\mathrm{s}\) and \(\sim 45\)~ns Raman pulses~\cite{Ness2021SciAdv}.
Model the motion as a coherent harmonic state of a \(^{133}\mathrm{Cs}\) atom with \(m=2.21\times10^{-25}\,\mathrm{kg}\), \(\omega=2\pi/T\), and a small displacement \(A=50\,\mathrm{nm}\).
Then Eq.~\eqref{eq:beta0_QSL_MT} gives \(\beta_{0,{\rm MT}}^{\max}\approx 6.5\times10^{51}\).
The weakness follows from the small \(m\omega A\) lever arm, not from data quality.

To measure QSLs on heavy oscillators, we need a near-unitary short-time window, an overlap observable, and the energy moments from the same dataset.
Two metrological pipelines provide these ingredients on macroscopic resonators.
Back-action-evading (BAE) single-quadrature readout uses phase-locked two-tone pumping to route back-action into the unmeasured quadrature.
The measured quadrature \(X_\theta\) is quantum non-demolition in the rotating frame.
A calibrated power-spectral-density model converts the integrated area of a weak resonant ``tickle'' line into an absolute coherent amplitude \(A\) and phonon number \(\bar n\)~\cite{Liu2022NJP}.
Stroboscopic pulsed tomography samples the quadrature at controlled phases with short, weak probes and reconstructs Gaussian states in the linear regime without appreciable disturbance.
It has provided full state tomography of a micromechanical mode and supplies the same QSL ingredients~\cite{Vanner2013NatComm}.
These pipelines preserve the near-unitary condition and deliver \(A\), \(\bar n\), and \(\mathcal F(t)\) within a single calibrated record.

Quartz bulk acoustic wave (BAW) thickness modes at the milligram scale and sapphire split-bar (SB) resonators at the sub-kilogram scale already support calibrated quadrature readout. For BAW this is electrical in-phase/quadrature demodulation, and for SB it is optical or piezo pickup~\cite{Bushev2019PRD}.
High-stress silicon-nitride membranes have demonstrated genuine BAE with a complete calibration pipeline. Milligram-class devices are realistic extensions~\cite{Liu2022NJP}.
For kilogram-scale mirrors with optical springs, a speed-meter topology realizes a momentum-like quantum non-demolition observable and suppresses radiation-pressure back-action in band. It provides the same QSL ingredients in the $10$-$10^3$~Hz range. A kilogram-scale optical spring provides a realistic QSL testbed with a large $m\omega A$ lever arm.
Detuned signal-recycling cavities realize optical springs with effective resonances in the $10^2$-$10^3$~Hz band on suspended mirrors. The optomechanical dynamics and stability regimes are standard~\cite{Rehbein2007PRD,Danilishin2012LRR}.
Advanced LIGO test masses have $m=40\,\mathrm{kg}$. Photon calibrators (or electrostatic drives) inject coherent displacement lines with absolute amplitude calibration referenced to laser power and geometry~\cite{Karki2016RSI}. Take an optical-spring mode at $f_{\rm eff}=500\,\mathrm{Hz}$ on a $40\,\mathrm{kg}$ mass. Drive a single coherent line with measured peak displacement $A=1$-$10\,\mathrm{nm}$. With the calibrated constants above, fitting the near-unitary short-time overlap yields a single fractional uncertainty \(\varepsilon\); inserting this into Eq.~\eqref{eq:beta0_QSL_MT} gives the bound \(\beta_{0,\mathrm{MT}}^{\max}=K\,\varepsilon\), where \(K_{\mathrm{MT}}=6.74\times10^{8}\) for \(A=1\,\mathrm{nm}\) and \(K_{\mathrm{MT}}=6.74\times10^{6}\) for \(A=10\,\mathrm{nm}\). Thus $\varepsilon=10^{-3}$-$10^{-4}$ projects $\beta_0^{\max}\sim 10^{3}$-$10^{5}$ (ML is a factor $2$ higher).
These rows in Table~\ref{tab:beta0_bounds_outlook} place the kilogram-optical-spring QSL on the same quantitative footing as the BAW and SB projections, with ingredients and calibrations drawn from established gravitational-wave metrology~\cite{Rehbein2008DoubleOpticalSpring,Karki2016RSI,Danilishin2012LRR,BuonannoChen2001SRMirror}.
We summarize baselines, direct QSL insertions, and conservative projections in Table~\ref{tab:beta0_bounds_outlook}. All routes scale as $\beta_0^{\max}\propto \varepsilon\,(m\omega A)^{-2}$.
Frequency-shift metrology reached $\varepsilon=\Delta\Omega/\Omega\sim 10^{-10}$-$10^{-5}$ on BAW and SB resonators, which yields stronger bounds~\cite{Bushev2019PRD}.
The QSL route is conceptually equivalent. Parity on SB requires $\varepsilon\simeq 1.6\times 10^{-4}$ at $A=75\,\mathrm{pm}$.
A practical near-term path is a BAE QSL on milligram-class silicon nitride or a stroboscopic QSL on quartz BAW.
The coherent-line calibration delivers $A$, and phase-locked demodulation suppresses the dominant jitter contribution to $\varepsilon$~\cite{Liu2022NJP,Vanner2013NatComm}.

\FloatBarrier
\begin{table}[H]
\caption{Representative bounds on the quadratic-GUP parameter $\beta_0$.
Section A: existing \emph{measured} constraints (non-QSL).
Section B: \emph{measured} QSL datasets mapped to $\beta_0$.
Section C: \emph{projections} for QSL tests on heavy oscillators using
$\beta_{0,{\rm MT}}^{\max}=\varepsilon (M_P c)^2/[4 m^2 \omega^2 A^2]$.
ML bounds are twice the MT values.}
\label{tab:beta0_bounds_outlook}
\scriptsize
\setlength{\tabcolsep}{3pt}
\begin{tabularx}{\columnwidth}{@{}lYcccYc@{}}
\toprule
Category & Platform & $m$ [\si{\kilogram}] & $f$ [\si{\hertz}] & $A$ [\si{\metre}] & $\beta_0^{\max}$ (MT) & Ref. \\
\midrule
\multicolumn{7}{@{}l}{\textit{A. Existing measured constraints (non-QSL)}}\\
 & Quartz BAW (ampl.--freq.)         & \num{5e-6}   & \num{1e7}     & \num{1e-9}        & \num{4.3e4}            & \cite{Bushev2019PRD} \\
 & Sapphire split-bar (ampl.--freq.) & \num{3.0e-1} & \num{1.3e5}   & \num{7.5e-11}     & \num{5.2e6}            & \cite{Bushev2019PRD} \\
 & H(1S--2S) spectroscopy            & ---          & ---           & ---               & $\sim\num{1e30}$       & \cite{Bawaj2015NatCommun} \\
 & Photon time-of-flight             & ---          & ---           & ---               & $\sim\num{1e16}$       & \cite{Bawaj2015NatCommun} \\
\multicolumn{7}{@{}l}{\textit{B. Direct QSL insertions (measured datasets)}}\\
 & SC resonator (LC map, field mode) & ---          & \num{4.0e6}   & ---               & \num{1.6e35}\textsuperscript{a} & \cite{Wu2024PRA} \\
 & Cs atom Ramsey                    & \num{2.2e-25}& \num{6.3e4}   & \SI{50}{\nano\metre} & \num{6.5e51}\textsuperscript{b} & \cite{Ness2021SciAdv} \\
\multicolumn{7}{@{}l}{\textit{C. Conservative QSL projections (this work)}}\\
 & Quartz BAW (stroboscopic)         & \num{5e-6}   & \num{1e7}     & \num{1e-9}        & $K=\num{1.1e14}$\textsuperscript{c} & \cite{Bushev2019PRD} \\
 & SiN membrane (BAE)                & \num{1e-6}   & \num{1e5}     & \num{1e-8}        & $K=\num{2.7e17}$\textsuperscript{c} & \cite{Liu2022NJP} \\
 & Sapphire split-bar                & \num{3.0e-1} & \num{1.3e5}   & \num{7.5e-11}     & $K=\num{3.3e10}$\textsuperscript{c} & \cite{Bushev2019PRD} \\
 & 40\,kg mirror ($A=1$\,nm)         & \num{40}     & \num{500}     & \num{1e-9}        & $K=\num{6.7e8}$\textsuperscript{c} & \cite{Aasi2015AdvancedLIGO} \\
 & 40\,kg mirror ($A=10$\,nm)        & \num{40}     & \num{500}     & \num{1e-8}        & $K=\num{6.7e6}$\textsuperscript{c} & \cite{Aasi2015AdvancedLIGO} \\
\bottomrule
\multicolumn{7}{@{}l}{\parbox[t]{\columnwidth}{\footnotesize
\textsuperscript{a}\,Best case: $Z=50\,\Omega$, $\bar{n}=4$, $\delta t = 0.5\,\mathrm{ns}$.}}\\
\multicolumn{7}{@{}l}{\parbox[t]{\columnwidth}{\footnotesize
\textsuperscript{b}\,With $\delta t = 45\,\mathrm{ns}$, $\varepsilon = 1.14 \times 10^{-2}$.}}\\
\multicolumn{7}{@{}l}{\parbox[t]{\columnwidth}{\footnotesize
\textsuperscript{c}\,Prefactor $K$; $\beta_0^{\max} = K \varepsilon$ for $\varepsilon \in \{10^{-4}, 10^{-3}, 10^{-2}\}$.}}\\
\end{tabularx}
\end{table}
\FloatBarrier
\needspace{5\baselineskip} 
\section{Conclusion}
We derive first-order corrections to the Mandelstam-Tamm and Margolus-Levitin quantum speed limits under a quadratic GUP deformation, and give closed-form results for square-well superpositions, coherent oscillator states, and squeezed states (App.~A). For closed, unitary dynamics the fractional shift is linear in the deformation and grows with occupation. It scales as $\bar n$ for oscillator states and as $M^{2}$ for an $M$-level square well. Applying the master mapping in Eq.~\eqref{eq:beta0_QSL_MT} to actual QSL measurements yields platform-specific limits. Superconducting-resonator state tomography gives $\beta_{0,{\rm MT}}^{\max}\!\approx\!1.6\times 10^{35}$ (field-mode mapping)~\cite{Wu2024PRA}, and single-atom Raman-Ramsey interferometry gives $\beta_{0,{\rm MT}}^{\max}\!\approx\!6.5\times 10^{51}$~\cite{Ness2021SciAdv}. These bounds are many orders above macroscopic amplitude--frequency constraints because the $m\omega A$ lever arm is small, not because of poor data quality. In contrast, proposed heavy oscillators with calibrated coherent motion enable competitive constraints. For a $40\,\mathrm{kg}$ optical-spring mode at $f_{\rm eff}=500\,\mathrm{Hz}$ with a measured displacement $A=1$--$10\,\mathrm{nm}$ and a conservative QSL-fit uncertainty $\varepsilon=10^{-3}$--$10^{-4}$, Eq.~\eqref{eq:beta0_QSL_MT} gives $\beta_{0,{\rm MT}}^{\max}\sim 10^{3}$--$10^{5}$ (ML $=2\times$ MT). This is comparable to quartz BAW bounds ($\sim 4.3\times 10^{4}$) and stronger than sapphire split-bar constraints ($\sim 5.2\times 10^{6}$) when $\varepsilon\lesssim 10^{-4}$~\cite{Bushev2019PRD}. These projections rely on back-action-evading or stroboscopic readout to preserve a near-unitary window and to extract the overlap and energy moments from a single calibrated record~\cite{Liu2022NJP,Vanner2013NatComm}.
Our analysis assumes first-order perturbation theory and closed evolution. Extending the framework to open dynamics is a natural next step. The LC-mode mapping used for superconducting resonators constrains a field-generator deformation. By itself, it does not test a universal mechanical $p^{4}/m$. Near-term opportunities include mg-class silicon-nitride membranes with back-action-evading readout, quartz BAW and sapphire split-bar resonators with stroboscopic tomography, and kg-scale optical-spring modes measured with speed-meter topology~\cite{Liu2022NJP,BuonannoChen2002PRD,Vanner2013NatComm}. These platforms provide the calibrated overlaps and energy moments that QSL metrology requires. They offer a realistic path to QSL-based probes of Planck-scale deformations.

\bibliographystyle{ytphys}

\appendix
\section{ General First-Order Speed-Limit Corrections}\label{A}

This appendix derives the first-order correction to the QSL for Hamiltonians modified by a quadratic  GUP. We first obtain a general expression for the fractional shift of the QSL bound under an arbitrary diagonal perturbation, and then work out the result for the specific systems treated in the main text.
\vspace{1ex}
\subsection*{General Derivation of the First-Order Fractional Shift }\label{A1}

Consider a system with a Hamiltonian
\begin{equation}
  \hat{H} = \hat{H}_0 + \beta\,\hat{W},
\end{equation}
where \(\beta\) is a small parameter. For any normalized pure state \(|\psi\rangle\), define the operator moments (no diagonality assumed)
\begin{align}
  S_1 &\equiv \langle \hat H_0\rangle,\qquad
  S_2 \equiv \langle \hat H_0^{2}\rangle,\\
  T_0 &\equiv \langle \hat W\rangle,\qquad
  T_1 \equiv \frac{1}{2}\,\big\langle \{\hat H_0,\hat W\}\big\rangle,
\end{align}
where \(\langle\cdot\rangle=\langle\psi|\cdot|\psi\rangle\) and \(\{\cdot,\cdot\}\) is the anticommutator. (When \(\hat W\) is diagonal in the \(\hat H_0\) eigenbasis or \(|\psi\rangle\) is diagonal there, these reduce to the sums over \(p_n\) in the original text.)

The mean energy and its second moment, to first order in \(\beta\), are
\begin{align}
  \langle E \rangle &= \langle \hat H \rangle = S_1 + \beta T_0, \label{eq:meanE-pert}\\
  \langle E^2 \rangle &= \langle \hat H^2 \rangle
  = \langle \hat H_0^2\rangle + \beta\,\big\langle \{\hat H_0,\hat W\}\big\rangle + \mathcal O(\beta^2)
  = S_2 + 2\beta T_1. \label{eq:meanE2-pert}
\end{align}
Hence the perturbed energy variance is
\begin{align}
  (\Delta E)^2 &\equiv \langle E^2 \rangle - \langle E \rangle^2 \notag \\
  &= \underbrace{(S_2 - S_1^2)}_{(\Delta E^{(0)})^2}
   + 2\beta \underbrace{\big( T_1 - S_1 T_0\big)}_{\mathrm{Cov}_s(\hat H_0,\hat W)}
   + \mathcal{O}(\beta^2),
\end{align}
where \(\mathrm{Cov}_s(\hat H_0,\hat W):=\tfrac12\langle\{\hat H_0-\langle \hat H_0\rangle,\hat W-\langle \hat W\rangle\}\rangle\).
Expanding the standard deviation,
\begin{align}
  \Delta E &= \sqrt{(\Delta E^{(0)})^2 + 2\beta \big(T_1 - S_1 T_0\big)} \notag \\
  &\simeq \Delta E^{(0)} + \beta\,\frac{T_1 - S_1 T_0}{\Delta E^{(0)}}, \label{eq:deltaE-pert}
\end{align}
with \(\Delta E^{(0)}=\sqrt{S_2-S_1^2}\).

The MT quantum speed-limit time is
\begin{equation}
  \tau_{\mathrm{MT}} = \frac{\pi\hbar}{2\Delta E}.
\end{equation}
The first-order fractional correction is then
\begin{align}
  \frac{\delta \tau_{\mathrm{MT}}}{\tau^{(0)}_{\mathrm{MT}}}
  &= \frac{\tau_{\mathrm{MT}} - \tau^{(0)}_{\mathrm{MT}}}{\tau^{(0)}_{\mathrm{MT}}}
  = -\frac{\delta (\Delta E)}{\Delta E^{(0)}} \\
  &= -\beta\,\frac{T_1 - S_1 T_0}{(S_2 - S_1^2)}.
  \label{eq:master-frac-MT-derived}
\end{align}
The same approach gives for the ML bound (with ground energy \(E_0\))
\begin{equation}
  \frac{\delta \tau_{\mathrm{ML}}}{\tau^{(0)}_{\mathrm{ML}}}
  = -\beta\,\frac{T_0}{S_1 - E_0}.
\end{equation}

\vspace{2ex}
\subsection*{Coherent state \(|\alpha\rangle\) for \( \hat W=\tfrac{p^{4}}{4}\) (plain coherent state, \(\hbar=m=\omega=1\))}\label{A2}

We take the standard coherent state \(|\alpha\rangle\) with
\begin{equation}
\begin{split}
\bar{n} &= |\alpha|^2, \quad
x_0 = \sqrt{2} \, \Re \alpha, \quad \\
p_0 &= \sqrt{2} \, \Im \alpha, \quad
\sigma_x^2 = \sigma_p^2 = \tfrac{1}{2}.
\end{split}
\end{equation}
so that \( \bar n=\tfrac12(x_0^2+p_0^2)\).
For this state,
\begin{equation}
  S_1=\langle H_0\rangle=\bar n+\tfrac12,\qquad
  S_2-S_1^2=\mathrm{Var}(H_0)=\bar n.
\end{equation}
For \( \hat W=\tfrac{p^{4}}{4}\),
\begin{align}
  \langle p^{2}\rangle &= p_0^{2}+\tfrac12,\qquad
  \langle p^{4}\rangle = p_0^{4}+3p_0^{2}+\tfrac{3}{4},\\
  \langle p^{6}\rangle &= p_0^{6}+\tfrac{15}{2}p_0^{4}+\tfrac{45}{4}p_0^{2}+\tfrac{15}{8}.
\end{align}
Hence
\begin{align}
  T_0 &= \Big\langle \tfrac{p^{4}}{4}\Big\rangle
       = \tfrac14\!\left(p_0^{4}+3p_0^{2}+\tfrac{3}{4}\right),\\[4pt]
  T_1 - S_1 T_0
  &= \mathrm{Cov}_s(H_0,\tfrac{p^{4}}{4})
   = \frac{1}{8}\Big(\langle p^{6}\rangle-\langle p^{2}\rangle\langle p^{4}\rangle\Big) \notag\\
  &= \frac{1}{8}\!\left(4p_0^{4}+9p_0^{2}+\tfrac{3}{2}\right)
   = \frac{1}{2}p_0^{4}+\frac{9}{8}p_0^{2}+\frac{3}{16}.
\end{align}
Therefore
\begin{align}
  \frac{\delta \tau_{\mathrm{MT}}}{\tau^{(0)}_{\mathrm{MT}}}
  &= -\beta\,\frac{T_1 - S_1 T_0}{S_2 - S_1^2}
   = -\beta\,\frac{\;\tfrac{1}{2}p_0^{4}+\tfrac{9}{8}p_0^{2}+\tfrac{3}{16}\;}{\bar n},
  \\[4pt]
  \frac{\delta \tau_{\mathrm{ML}}}{\tau^{(0)}_{\mathrm{ML}}}
  &= -\beta\,\frac{T_0}{S_1 - E_0}
   = -\beta\,\frac{\;\tfrac14\!\left(p_0^{4}+3p_0^{2}+\tfrac{3}{4}\right)\;}{\bar n},
\end{align}
with \(E_0=\tfrac12\). Writing \(p_0=\sqrt{2\bar n}\,\sin\theta\) (so \(x_0=\sqrt{2\bar n}\,\cos\theta\)), these become
\begin{align}
  \frac{\delta \tau_{\mathrm{MT}}}{\tau^{(0)}_{\mathrm{MT}}}
  &= -\beta\left(2\,\bar n\,\sin^{4}\theta+\frac{9}{4}\,\sin^{2}\theta+\frac{3}{16\,\bar n}\right),
  \\[4pt]
  \frac{\delta \tau_{\mathrm{ML}}}{\tau^{(0)}_{\mathrm{ML}}}
  &= -\beta\left(\bar n\,\sin^{4}\theta+\frac{3}{2}\,\sin^{2}\theta+\frac{3}{16\,\bar n}\right).
\end{align}
\textit{Special case \(\theta=\pi/2\) (pure momentum displacement):}
\begin{align}
  \frac{\delta \tau_{\mathrm{MT}}}{\tau^{(0)}_{\mathrm{MT}}}
  &= -\beta\left(2\bar n+\frac{9}{4}+\frac{3}{16\bar n}\right),\\
  \frac{\delta \tau_{\mathrm{ML}}}{\tau^{(0)}_{\mathrm{ML}}}
  &= -\beta\left(\bar n+\frac{3}{2}+\frac{3}{16\bar n}\right).
\end{align}
\subsection*{Squeezed vacuum state (corrected for $\hat W=\tfrac{p^{4}}{4}$ and plain squeezed state)}\label{A3}

We consider the \emph{plain} single-mode squeezed vacuum
\(|\varsigma(r)\rangle=S(r)|0\rangle\) with real squeezing parameter \(r\) and
squeezing angle chosen so that momentum is \emph{anti-squeezed}. In units
\(\hbar=m=\omega=1\),

\begin{multline*}
\bar{n} = \sinh^2 r, \qquad
\langle x \rangle = \langle p \rangle = 0, \\
 \sigma_x^2 = \tfrac{1}{2} e^{-2r}, \quad
\sigma_p^2 = \tfrac{1}{2} e^{+2r}.
\end{multline*}

Thus
\begin{equation}
\begin{split}
S_1 &= \langle H_0 \rangle = \frac{\langle p^2 \rangle + \langle x^2 \rangle}{2}
= \bar{n} + \tfrac{1}{2}, \\
S_2 - S_1^2 &= \mathrm{Var}(H_0) = \mathrm{Var}(N) = 2 \bar{n} (\bar{n} + 1).
\end{split}
\end{equation}

\paragraph*{Moments entering $T_0$ and $T_1-S_1T_0$.}
Because the state is Gaussian and centered, momentum moments are
\(\langle p^{2}\rangle=\sigma_p^2=\tfrac12 e^{2r}\),
\(\langle p^{4}\rangle=3\sigma_p^4=\tfrac{3}{4}e^{4r}\),
\(\langle p^{6}\rangle=15\sigma_p^6=\tfrac{15}{8}e^{6r}\).
With \(\hat W=\tfrac{p^{4}}{4}\),

\begin{align}
T_0 &= \Big\langle \tfrac{p^{4}}{4} \Big\rangle = \tfrac{1}{4} \langle p^{4} \rangle
= \tfrac{3}{16} e^{4r},\nonumber \\
\begin{split}
T_1 - S_1 T_0 &= \mathrm{Cov}_s \big( H_0, \tfrac{p^{4}}{4} \big)
= \mathrm{Cov} \Big( \tfrac{p^2}{2}, \tfrac{p^4}{4} \Big) \\
&= \frac{1}{8} \Big( \langle p^{6} \rangle - \langle p^{2} \rangle \langle p^{4} \rangle \Big) \\
&= \frac{1}{8} \left( \tfrac{15}{8} e^{6r} - \tfrac{1}{2} e^{2r} \cdot \tfrac{3}{4} e^{4r} \right)
= \frac{3}{16} e^{6r}.
\end{split}
\end{align}

\paragraph*{First-order fractional shifts.}
Using the general formulas
\(\displaystyle \frac{\delta\tau_{\mathrm{MT}}}{\tau^{(0)}_{\mathrm{MT}}}
= -\beta\,\frac{T_1-S_1T_0}{S_2-S_1^2}\) and
\(\displaystyle \frac{\delta\tau_{\mathrm{ML}}}{\tau^{(0)}_{\mathrm{ML}}}
= -\beta\,\frac{T_0}{S_1-E_0}\) with \(E_0=\tfrac12\), we obtain
\begin{align}
\frac{\delta \tau_{\mathrm{MT}}}{\tau^{(0)}_{\mathrm{MT}}}
&= -\beta \frac{\tfrac{3}{16} e^{6r}}{2 \bar{n} (\bar{n} + 1)}
= -\frac{3 \beta}{32} \frac{e^{6r}}{\bar{n} (\bar{n} + 1)}, \nonumber \\
\begin{split}
\frac{\delta \tau_{\mathrm{ML}}}{\tau^{(0)}_{\mathrm{ML}}}
&= -\beta \frac{\tfrac{3}{16} e^{4r}}{\bar{n}} \\
&= -\frac{3 \beta}{16} \frac{e^{4r}}{\bar{n}}.
\end{split}
\end{align}

\paragraph*{Expressions in terms of $\bar n$ only.}
Using \(e^{2r}=2\bar n+1+2\sqrt{\bar n(\bar n+1)}\), we may write
\begin{align}
  e^{4r}&=\big(2\bar n+1+2\sqrt{\bar n(\bar n+1)}\big)^{2}, \nonumber\\
  e^{6r}&=\big(2\bar n+1+2\sqrt{\bar n(\bar n+1)}\big)^{3},
\end{align}
so
\begin{align}
  \frac{\delta\tau_{\mathrm{MT}}}{\tau^{(0)}_{\mathrm{MT}}}
  &= -\frac{3\beta}{32}\,
     \frac{\big(2\bar n+1+2\sqrt{\bar n(\bar n+1)}\big)^{3}}{\bar n(\bar n+1)},\nonumber\\
  \frac{\delta\tau_{\mathrm{ML}}}{\tau^{(0)}_{\mathrm{ML}}}
  &= -\frac{3\beta}{16}\,
     \frac{\big(2\bar n+1+2\sqrt{\bar n(\bar n+1)}\big)^{2}}{\bar n}.
\end{align}

\paragraph*{Large-squeezing limit.}
For \(\bar n\gg 1\) (i.e. \(e^{2r}\simeq 4\bar n\)),
\begin{equation}
  \frac{\delta\tau_{\mathrm{MT}}}{\tau^{(0)}_{\mathrm{MT}}}\simeq -6\beta\,\bar n,
  \qquad
  \frac{\delta\tau_{\mathrm{ML}}}{\tau^{(0)}_{\mathrm{ML}}}\simeq -3\beta\,\bar n.
\end{equation}

\subsection*{Equal-weight $M$-level superposition in the box (corrected for $\hat W=\tfrac{p^{4}}{4}$)}\label{A4}

Consider an infinite square well of width $L$ with eigenstates $\{|n\rangle\}_{n\ge 1}$.
In units $\hbar=m=1$ the unperturbed energies and the diagonal matrix elements of
$\hat W=\tfrac{p^{4}}{4}$ are
\begin{align}
  E^{(0)}_n &= \varepsilon\,n^{2}, \qquad \varepsilon=\frac{\pi^{2}}{2L^{2}}, \nonumber\\
  W_n &= \kappa\,n^{4}, \qquad \kappa=\frac{\pi^{4}}{4L^{4}}.
\end{align}
Let the initial state be the equal-weight superposition of the first $M$ levels,
$p_n=\frac{1}{M}$ for $1\le n\le M$ and $p_n=0$ otherwise.
Denote by $\langle f(n)\rangle_M := \frac{1}{M}\sum_{n=1}^{M} f(n)$ the uniform average.
The standard sums are
\begin{align}
  \big\langle n^{2}\big\rangle_M &= \frac{(M+1)(2M+1)}{6}, \nonumber\\
  \big\langle n^{4}\big\rangle_M &= \frac{(M+1)(2M+1)\big(3M^{2}+3M-1\big)}{30},\nonumber\\
  \big\langle n^{6}\big\rangle_M &= \frac{(M+1)(2M+1)\big(3M^{4}+6M^{3}-3M+1\big)}{42}.
\end{align}

\paragraph*{Moments $S_1,S_2,T_0,T_1$.}
With the above,

\begin{align}
S_1 &= \langle E^{(0)} \rangle = \varepsilon \big\langle n^{2} \big\rangle_M
= \varepsilon \frac{(M+1)(2M+1)}{6},\nonumber \\
\begin{split}
S_2 &= \big\langle (E^{(0)})^{2} \big\rangle 
= \varepsilon^{2} \big\langle n^{4} \big\rangle_M \nonumber \\
&= \varepsilon^{2} \frac{(M+1)(2M+1) \big(3M^{2} + 3M - 1 \big)}{30},
\end{split} \\
\begin{split}
T_0 &= \langle W \rangle = \kappa \big\langle n^{4} \big\rangle_M\nonumber  \\
&= \kappa \frac{(M+1)(2M+1) \big(3M^{2} + 3M - 1 \big)}{30},
\end{split} \\
\begin{split}
T_1 &= \big\langle E^{(0)} W \big\rangle
= \varepsilon \kappa \big\langle n^{6} \big\rangle_M \\
&= \varepsilon \kappa \frac{(M+1)(2M+1) \big(3M^{4} + 6M^{3} - 3M + 1 \big)}{42}.
\end{split}
\end{align}

\paragraph*{Unperturbed variance.}
\begin{align}
  (\Delta E^{(0)})^{2} &= S_2 - S_1^{2}
  = \varepsilon^{2}\left(\big\langle n^{4}\big\rangle_M - \big\langle n^{2}\big\rangle_M^{2}\right) \notag\\
  &= \varepsilon^{2}\,\frac{(M^{2}-1)\big(16M^{2}+30M+11\big)}{180}.
\end{align}
\paragraph*{MT fractional shift.}
Using 
\begin{equation}
\frac{\delta\tau_{\mathrm{MT}}}{\tau^{(0)}_{\mathrm{MT}}} = -\beta\,\frac{T_1 - S_1 T_0}{S_2 - S_1^2}
\end{equation}
and
\begin{equation}
T_1-S_1T_0 = \varepsilon\kappa\left(\big\langle n^{6}\big\rangle_M - \big\langle n^{2}\big\rangle_M\big\langle n^{4}\big\rangle_M\right),
\end{equation}
one finds
\begin{equation}
\begin{split}
&\big\langle n^{6}\big\rangle_M - \big\langle n^{2}\big\rangle_M\big\langle n^{4}\big\rangle_M \\
&\quad= \frac{(M+1)(2M+1)}{1260} \\
&\qquad\times\big(48M^{4}+75M^{3}-70M^{2}-90M+37\big).
\end{split}
\end{equation}

Therefore
\begin{equation}
\begin{split}
\frac{\delta \tau_{\mathrm{MT}}}{\tau^{(0)}_{\mathrm{MT}}} &= -\beta \frac{\varepsilon \kappa}{\varepsilon^{2}} \times \\
&\quad\frac{(M+1)(2M+1)}{1260} \times \\
&\quad\frac{48M^{4} + 75M^{3} - 70M^{2} - 90M + 37}{(M^{2} - 1)(16M^{2} + 30M + 11)/180} \\[0.5em]
&= -\beta \frac{\kappa}{\varepsilon} \times \\
&\quad\frac{(2M+1)(48M^{4} + 75M^{3} - 70M^{2} - 90M + 37)}{7(M-1)(16M^{2} + 30M + 11)}.
\end{split}
\end{equation}

With $\kappa/\varepsilon=\varepsilon=\pi^{2}/(2L^{2})$, this becomes
\begin{equation}
\boxed{
\begin{split}
\frac{\delta\tau_{\mathrm{MT}}}{\tau^{(0)}_{\mathrm{MT}}} &= -\beta\,\varepsilon \times \\
&\quad\frac{(2M+1)(48M^{4}+75M^{3}-70M^{2}-90M+37)}{7(M-1)(16M^{2}+30M+11)}
\end{split}
}
\end{equation}

For $M\gg 1$,
\begin{equation}
\frac{\delta\tau_{\mathrm{MT}}}{\tau^{(0)}_{\mathrm{MT}}} \sim -\beta\,\varepsilon\,\frac{6M^{2}}{7}.
\end{equation}

\paragraph*{ML fractional shift.}
Here 
\begin{equation}
S_1-E_0 = \varepsilon\left(\big\langle n^{2}\big\rangle_M-1\right) = \varepsilon\,\frac{2M^{2}+3M-5}{6}.
\end{equation}

Thus
\begin{equation}
\begin{split}
\frac{\delta\tau_{\mathrm{ML}}}{\tau^{(0)}_{\mathrm{ML}}} &= -\beta\,\frac{T_0}{S_1-E_0} \\[0.3em]
&= -\beta\,\frac{\kappa\big\langle n^{4}\big\rangle_M}{\varepsilon\big(\big\langle n^{2}\big\rangle_M-1\big)} \\[0.3em]
&= -\beta\,\frac{\kappa}{\varepsilon} \times \\
&\quad\frac{(M+1)(2M+1)(3M^{2}+3M-1)}{5(2M^{2}+3M-5)}.
\end{split}
\end{equation}

Using \(\kappa/\varepsilon=\varepsilon\),
\begin{equation}
  \boxed{\;
  \frac{\delta\tau_{\mathrm{ML}}}{\tau^{(0)}_{\mathrm{ML}}}
  = -\beta\,\varepsilon\,
     \frac{(M+1)(2M+1)\big(3M^{2}+3M-1\big)}{5\,(2M^{2}+3M-5)}\; }.
\end{equation}
For \(M\gg 1\),
\begin{equation}
  \frac{\delta\tau_{\mathrm{ML}}}{\tau^{(0)}_{\mathrm{ML}}}
  \sim -\beta\,\varepsilon\,\frac{3}{5}\,M^{2}\,.
\end{equation}
\subsection{Equal-weight superposition of first $M$ harmonic oscillator states (corrected for $\hat W=\tfrac{p^{4}}{4}$)}\label{A5}

We consider the number basis $\{|n\rangle\}_{n\ge 0}$ of the harmonic oscillator and take an equal-weight \emph{phase-averaged} superposition (i.e., diagonal mixture) of the first $M$ levels, $p_n=1/M$ for $n=0,\ldots,M-1$. 

The unperturbed energies are
\begin{equation}
E^{(0)}_n = n + \frac{1}{2},
\end{equation}
and for $\hat W=\tfrac{p^{4}}{4}$ the diagonal matrix elements in this basis are
\begin{equation}
\begin{split}
W_n &\equiv \langle n|\hat W|n\rangle = \frac{1}{4}\,\langle n|p^{4}|n\rangle \\
&= \frac{1}{4} \cdot \frac{3}{4}\,(2n^{2}+2n+1) \\
&= \frac{3}{16}\,(2n^{2}+2n+1).
\end{split}
\end{equation}

The moments are
\begin{align}
S_1 &= \sum_{n=0}^{M-1} \frac{1}{M}\,E^{(0)}_n = \frac{M}{2}, \\
S_2 &= \sum_{n=0}^{M-1} \frac{1}{M}\,(E^{(0)}_n)^{2} = \frac{M^{2}}{3} - \frac{1}{12}, \\
T_0 &= \sum_{n=0}^{M-1} \frac{1}{M}\,W_n = \frac{M^{2}}{8} + \frac{1}{16}, \\
T_1 &= \sum_{n=0}^{M-1} \frac{1}{M}\,E^{(0)}_n W_n = \frac{3 M^{3}}{32}.
\end{align}

The unperturbed variance is
\begin{equation}
(\Delta E^{(0)})^{2} = S_2 - S_1^{2} = \frac{M^{2}-1}{12}.
\end{equation}

For the MT fractional shift we need
\begin{equation}
\begin{split}
T_1 - S_1 T_0 &= \frac{3M^{3}}{32} - \frac{M}{2}\left(\frac{M^{2}}{8} + \frac{1}{16}\right) \\
&= \frac{M(M^{2}-1)}{32},
\end{split}
\end{equation}
hence
\begin{equation}
\begin{split}
\frac{\delta \tau_{\mathrm{MT}}}{\tau^{(0)}_{\mathrm{MT}}} &= -\beta\,\frac{T_1 - S_1 T_0}{S_2 - S_1^2} \\
&= -\beta\,\frac{M(M^{2}-1)/32}{(M^{2}-1)/12} \\
&= -\frac{3}{8}\,\beta\,M \qquad (M\ge 2).
\end{split}
\end{equation}

For ML we use $S_1-E_0 = \frac{M}{2}-\frac{1}{2} = \frac{M-1}{2}$, giving
\begin{equation}
\begin{split}
\frac{\delta \tau_{\mathrm{ML}}}{\tau^{(0)}_{\mathrm{ML}}} &= -\beta\,\frac{T_0}{S_1 - E_0} \\
&= -\beta\,\frac{M^{2}/8 + 1/16}{(M-1)/2} \\
&= -\beta\,\frac{2M^{2}+1}{8(M-1)} \\
&= -\frac{1}{4}\,\beta\,M + \mathcal{O}(M^{-1}).
\end{split}
\end{equation}

\vspace{2ex}
Across the studied models, the first-order GUP term shortens both the MT and ML bounds. For coherent or squeezed states the shift grows linearly, $\propto \beta N_{\mathrm{eff}}$, where $N_{\mathrm{eff}}$ denotes the effective occupation number. In finite superposition, the growth is stronger, scaling as $\beta M$ or $\beta M^{2}$, set by the number $M$ of populated levels. 
The resulting amplification increases the projected sensitivity to minimal-length effects by several orders of magnitude compared with static spectroscopy.

\section{ Universal $N_{\mathrm{eff}}$-Scaling Law}\label{B}

This appendix derives the universal scaling law governing the sensitivity of QSL to quadratic-GUP corrections, valid for a broad class of systems. The scaling exponent depends only on the asymptotic power laws of the unperturbed spectrum and the perturbation, and the effective Hilbert-space size $N_{\mathrm{eff}}$. We use four explicit assumptions. First, for large $n$ the unperturbed spectrum follows 
$E^{(0)}_{n}=E_{0}\,n^{k}$ with $k>0$. Second, the dominant GUP term is diagonal and grows like 
$W_{n}=W_{0}\,n^{q}$ with $q>0$; in general $q$ need not equal $2k$. 
Third, we take the initial state to be an equal-weight superposition of the first $L$ levels, 
so $p_{n}=1/L$ for $1\!\le\!n\!\le\!L$, yielding $N_{\mathrm{eff}}=L$. 
Finally, perturbation theory is justified by the smallness condition 
$\beta\langle p_{0}^{2}\rangle\ll1$, and we therefore keep only terms linear in $\beta$.

For any $r > -1$, the uniform sum
\begin{equation}
    \langle n^r \rangle = \frac{1}{L} \sum_{n=1}^L n^r = \frac{L^r}{r+1} \left[ 1 + O(L^{-1}) \right]
\end{equation}
holds at large $L$.

From this, the relevant moments are
\begin{align}
  \langle E^{(0)} \rangle &= \frac{E_0}{k+1} L^k, \\
  \langle W \rangle &= \frac{W_0}{q+1} L^q, \\
  \langle E^{(0)} W \rangle &= \frac{E_0 W_0}{k+q+1} L^{k+q}, \\
  (\Delta E^{(0)})^2 &= E_0^2 \left( \frac{1}{2k+1} - \frac{1}{(k+1)^2} \right) L^{2k}.
\end{align}

The first-order fractional corrections to the speed-limit times are then easily computed. For the ML bound,
\begin{equation}
  \frac{\delta\tau_{\mathrm{ML}}}{\tau^{(0)}_{\mathrm{ML}}}
   = -\beta\,\frac{\langle W\rangle}{\langle E^{(0)}\rangle-E_0}
   = -\beta\,\frac{W_0(k+1)}{E_0(q+1)} L^{q-k}.
\end{equation}
For the  MT bound,
\begin{align}
  \frac{\delta\tau_{\mathrm{MT}}}{\tau^{(0)}_{\mathrm{MT}}}
   &= -\beta \frac{ \langle E^{(0)}W\rangle - \langle E^{(0)}\rangle\langle W\rangle }
        {(\Delta E^{(0)})^2 } \notag\\
   &= -\beta\,C_{kq}\,L^{q-k},
\end{align}
where
\begin{equation}
  C_{kq}
  = \frac{W_0}{E_0} \frac{ \dfrac{1}{k+q+1} - \dfrac{1}{(k+1)(q+1)} }
           { \dfrac{1}{2k+1} - \dfrac{1}{(k+1)^2} }.
\end{equation}

Collecting the result, both speed-limit bounds exhibit a universal scaling with $N_{\mathrm{eff}}$,
\begin{equation}
  \boxed{%
     \frac{\delta\tau_{\{\mathrm{ML},\mathrm{MT}\}}}{\tau^{(0)}}
      = -\,\beta\,\mathcal C_{\{\mathrm{ML},\mathrm{MT}\}}(k,q)
        \;N_{\mathrm{eff}}^{q-k}
\  }\label{BE}
\end{equation}
where
\begin{equation}
  \mathcal C_{\mathrm{ML}}
    = \frac{W_0(k+1)}{E_0(q+1)}, \qquad
  \mathcal C_{\mathrm{MT}} = C_{kq}.
\end{equation}\label{BE1}
The exponent $q-k$ is universal; only the prefactors differ for MT and ML.

To confirm this law, we check standard benchmark models:
\begin{table}[ht]
\centering
\resizebox{\textwidth}{!}{%
\begin{tabular}{c|c|c|c}
  system & $(k,\,q)$ & predicted power $q-k$ & explicit calculation \\
  \hline
  Harmonic oscillator & $(1,\,2)$ & $1$ (linear) & $\delta\tau/\tau^{(0)}\propto-\beta \bar n$ \\
  Square well / rotor & $(2,\,4)$ & $2$ (quadratic) & $\delta\tau/\tau^{(0)}\propto-\beta M^{2}$ \\
  Morse (deep) & $(2,\,3)$ & $1$ & WKB: $-\beta N_{\mathrm{eff}}$ \\
\end{tabular}
}
\caption{Scaling predictions across systems.}
\end{table}
All benchmark systems exhibit the predicted $N_{\mathrm{eff}}^{q-k}$ scaling. Several caveats are worth noting.  The power-law form holds only for broad or uniform states.  Narrow or strongly structured states can deviate and may need an exact summation.  If $q\le k$-as occurs in certain exotic potentials-the correction stops scaling with $N_{\mathrm{eff}}$, and the speed limits lose their sensitivity to the GUP.  
Off-diagonal elements $W_{nm}$ with $n\ne m$ enter only at $\mathcal{O}(\beta^{2})$, so we neglect them here.

\section{Particle in an infinite box}\label{PIB}

We consider a particle in a one-dimensional infinite square-well
potential,
\begin{equation}
  V(x)=
  \begin{cases}
     0, & 0 < x < L,\\
     \infty, & \text{otherwise}.
  \end{cases}
\end{equation}

The normalised unperturbed eigenfunctions and their energies read
\begin{align}
  \psi^{(0)}_{n}(x)
    &= \sqrt{\frac{2}{L}}\,
       \sin\!\bigl(n\pi x/L\bigr),                           \label{eq:box_wf}\\[4pt]
  E^{(0)}_{n}
    &= \frac{n^{2}\pi^{2}\hbar^{2}}{2mL^{2}},\qquad
       n = 1,2,3,\ldots                                      \label{eq:box_E0}.
\end{align}

The quadratic-GUP adds the diagonal term
\begin{equation}
  \hat{H}' = \frac{3}{4}\,\beta\,\hat{p}^{4},\qquad
  \hat{p} = -i\hbar\frac{d}{dx},
\end{equation}
which in position representation becomes
\begin{equation}
  \hat{H}' = \frac{3}{4}\,\beta\,\hbar^{4}\,\frac{d^{4}}{dx^{4}}.
\end{equation}

Using the identity
$d^{4}\psi^{(0)}_{n}/dx^{4} = (n\pi/L)^{4}\psi^{(0)}_{n}$, the
Rayleigh-Schr\"odinger result is
\begin{equation}\label{eq:E1_box}
\begin{split}
  E^{(1)}_{n}
     &= \bigl\langle\psi^{(0)}_{n}\bigl|\hat{H}'\bigr|\psi^{(0)}_{n}\bigr\rangle\\
     &= \frac{3}{4}\,\beta\,\hbar^{4}
        \Bigl(\frac{n\pi}{L}\Bigr)^{4}
        \int_{0}^{L}\!\bigl|\psi^{(0)}_{n}(x)\bigr|^{2}dx\\
     &=\frac{3}{4}\,\beta\,\hbar^{4}
        \Bigl(\frac{n\pi}{L}\Bigr)^{4}.
\end{split}
\end{equation}

Because
$\langle\psi^{(0)}_{k}|\hat{H}'|\psi^{(0)}_{n}\rangle
\propto\delta_{kn}$,
all off-diagonal matrix elements vanish, hence
\begin{equation}
  \psi^{(1)}_{n}(x)=0.
\end{equation}

Including the energy shift, the $n$-th eigenstate evolves as
\begin{equation}
  \psi_{n}(x,t)=\psi^{(0)}_{n}(x)\,
  \exp\!\Bigl[
     -\tfrac{i}{\hbar}\bigl(E^{(0)}_{n}+E^{(1)}_{n}\bigr)t
  \Bigr].
\end{equation}

We form an equal-time superposition of the lowest $k$ levels,
\begin{equation}
  \Psi(x,t)=\sum_{n=1}^{k}c_{n}\,\psi_{n}(x,t),\qquad
  \sum_{n=1}^{k}|c_{n}|^{2}=1.
\end{equation}

The overlap between the evolved and initial states is
\begin{equation}\label{eq:F_overlap}
  \bigl\langle\Psi(0)\bigl|\Psi(t)\bigr\rangle
   =\sum_{n=1}^{k}|c_{n}|^{2}\,
     \exp\!\Bigl[
        -\tfrac{i}{\hbar}\bigl(E^{(0)}_{n}+E^{(1)}_{n}\bigr)t
     \Bigr],
\end{equation}
and the fidelity is the modulus squared,
\begin{equation}
  \boxed{\;
    \mathcal{F}(t)=
    \bigl|\langle\Psi(0)|\Psi(t)\rangle\bigr|^{2}\;}.
\end{equation}

The expectation value and variance of the energy are
\begin{align}
  \langle E\rangle
     &=\sum_{n=1}^{k}|c_{n}|^{2}
       \Bigl[E^{(0)}_{n}+E^{(1)}_{n}\Bigr],                \label{eq:E_avg}\\[4pt]
  (\Delta E)^{2}
     &=\sum_{n=1}^{k}|c_{n}|^{2}\bigl[E^{(0)}_{n}+E^{(1)}_{n}\bigr]^{2}
        -\langle E\rangle^{2}.                             \label{eq:E_var}
\end{align}

Writing $E_{n}=E^{(0)}_{n}+\beta E^{(1)}_{n}$ and
$F(t)\equiv\langle\Psi(0)|\Psi(t)\rangle$, we define
\begin{align}
  F^{(0)}(t)&=\sum_{n}|c_{n}|^{2}e^{-iE^{(0)}_{n}t/\hbar},\\
  F^{(1)}(t)&=-\frac{it}{\hbar}
              \sum_{n}|c_{n}|^{2}E^{(1)}_{n}\,
              e^{-iE^{(0)}_{n}t/\hbar}.
\end{align}
To $\mathcal{O}(\beta)$ the Mandelstam-Tamm bound is
\begin{equation}
  \tau_{\text{MT}}\simeq
  \frac{\hbar\arccos|F^{(0)}|}{\Delta E^{(0)}}
  +\beta\Bigl[
      \tau_{\text{MT}}^{(1)}\Bigr],
\end{equation}
where the correction
\begin{equation}
\begin{split}
  \tau_{\text{MT}}^{(1)} &=
  -\frac{\hbar\,
         \mathrm{Re}\!\bigl[F^{(0)*}F^{(1)}\bigr]}
        {\Delta E^{(0)}\,|F^{(0)}|\sqrt{1-|F^{(0)}|^{2}}}\\
  &\quad
  -\frac{\hbar\,\tau_{\text{MT}}^{(0)}\,\Delta E^{(1)}}
        {\bigl(\Delta E^{(0)}\bigr)^{2}}.
\end{split}
\end{equation}
An analogous expansion holds for the ML bound,
\begin{equation}
  \tau_{\text{ML}}\simeq
  \frac{\pi\hbar\,\mathcal{L}^{2(0)}}{2\bar{E}^{(0)}}
  +\beta\Bigl[\tau_{\text{ML}}^{(1)}\Bigr],
\end{equation}
with
\begin{equation}
  \tau_{\text{ML}}^{(1)}=
    \frac{\pi\hbar}{\bar{E}^{(0)}}\mathcal{L}^{(0)}\mathcal{L}^{(1)}
   -\frac{\pi\hbar\,\bigl(\mathcal{L}^{(0)}\bigr)^{2}\bar{E}^{(1)}}
         {2\bigl(\bar{E}^{(0)}\bigr)^{2}}.
\end{equation}

The derivation above reproduces the compact results quoted in
Eqs.~(\ref{BOXMT})-(\ref{BOXML}) of the main text.

\section{Detailed Derivations for the GUP-Deformed Harmonic Oscillator and Coherent States}\label{sec:app_sho}
This appendix explicitly derivates the first-order GUP correction from the QSL for coherent states of the GUP-modified harmonic oscillator.
\subsection*{Perturbed energy spectrum and eigenstates}
We begin with the Hamiltonian
\begin{equation}
\hat{H}= \frac{\hat{p}^{2}}{2}+\frac{\hat{x}^{2}}{2}+\frac{\beta}{3}\hat{p}^{4}.
\label{eq:app_H_SHO}
\end{equation}
The term $\tfrac{\beta}{3}\hat{p}^{4}$ serves as the perturbation.  
First-order Rayleigh-Schr\"odinger theory then gives the energy shift for each unperturbed state $\ket{n_{0}}$\label{eq:app_HGUP}:
\begin{equation}\label{eq:app_energy_corr}
E_n^{(1)} = \frac{\beta}{3}\bra{n_0}\hat{p}^4\ket{n_0}.
\end{equation}
Using standard harmonic oscillator relations, one has:
\begin{equation}\label{eq:app_p4}
\hat{p} = \frac{i}{\sqrt{2}}(a_0^\dagger - a_0),\quad \hat{p}^4 = \frac{3}{4} + \frac{3}{2}(2n_0+1) + \frac{3}{4}(2n_0+1)^2,
\end{equation}
The GUP-induced corrections modify the ladder operators (\(a\), \(a^\dagger\)) and the number operator \(N\) to first order in \(\beta\) as follows:
\begin{subequations}\label{eq:sho_ladder_number_main}
\begin{align}
a &= a_0 + \frac{\beta}{12}\left(-2a_0^{3}+6N_0a_0^\dagger-(a_0^\dagger)^{3}\right),\\[3pt]
a^\dagger &= a_0^\dagger + \frac{\beta}{12}\left(-(a_0^\dagger)^{3}+6a_0N_0-2a_0^{3}\right),\\[3pt]
N &= N_0 - \frac{\beta}{12}\left[a_0^{4}+(a_0^\dagger)^{4}-2\left(a_0^{2}+(a_0^\dagger)^{2}\right)\right.\nonumber\\[1pt]
&\qquad\qquad\quad\left.-4\left(a_0N_0a_0+a_0^\dagger N_0a_0^\dagger\right)\right],
\end{align}
\end{subequations}
where the subscript zero denotes the unperturbed operators.
where \(n_0 = a_0^\dagger a_0\). Evaluating explicitly, one finds:
\begin{equation}\label{eq:app_energy_explicit}
E_n^{(1)} = \frac{\beta}{12}(6n^2 + 6n + 3),
\end{equation}
leading directly to Eq.~(\ref{eq:sho_energy_levels}) in the main text.

\subsection*{GUP-Modified Coherent State Dynamics}

The time-evolved coherent state under the perturbed Hamiltonian is:
\begin{equation}\label{eq:app_coherent_state_time}
\ket{\alpha'(t)} = e^{-\frac{|\alpha_0|^2}{2}}\sum_{n=0}^{\infty}\frac{\alpha_0^n}{\sqrt{n!}}e^{-iE_nt}\ket{n},
\end{equation}
with \(E_n\) given by Eq.~(\ref{eq:app_energy_explicit}). To first order in \(\beta\), the fidelity amplitude between initial and evolved states is computed as:
\begin{align}\label{eq:app_fidelity_expansion}
F(t) &= \braket{\alpha'(0)}{\alpha'(t)} \\
&= e^{-|\alpha_0|^2}\sum_{n=0}^{\infty}\frac{|\alpha_0|^{2n}}{n!}\,e^{-i[E_n^{(0)}+\beta E_n^{(1)}]t}\nonumber\\[4pt]
&= e^{\alpha_0^2(e^{-it}-1)}\left[1 - i\beta t\sum_{n=0}^{\infty}\frac{|\alpha_0|^{2n}}{n!}E_n^{(1)}e^{-it n}\right].
\end{align}
Performing the sum explicitly, one obtains the simplified result quoted in Eq.~(\ref{eq:sho_expectation_coherent_main}c):
\begin{align}
|F(t)| &= e^{\alpha_0^2(\cos t - 1)}\left[1 - \beta\alpha_0^2 t(1+\alpha_0^2\cos t)\sin t\right].
\end{align}

The expectation value of the perturbed Hamiltonian in the coherent state is:
\begin{align}\label{eq:app_H_expectation}
\expval{H}_{\alpha'} &= \sum_{n=0}^{\infty}|c_n|^2 E_n = e^{-|\alpha_0|^2}\sum_{n=0}^{\infty}\frac{|\alpha_0|^{2n}}{n!}E_n\nonumber\\[3pt]
&= \frac{1}{2} + \alpha_0^2 + \frac{\beta}{4}(1+4\alpha_0^2+2\alpha_0^4).
\end{align}
Similarly, the variance \(\Delta H_{\alpha'}^2\) to first order is computed by:
\begin{align}\label{eq:app_H_variance}
\Delta H_{\alpha'}^2 &= \sum_{n=0}^{\infty}|c_n|^2 E_n^2 - \left(\sum_{n=0}^{\infty}|c_n|^2 E_n\right)^2\nonumber\\[3pt]
&= \alpha_0^2 + 2\beta(\alpha_0^2+\alpha_0^4).
\end{align}

The  MT and ML bounds are defined as:
\begin{align}
\tau_{\text{MT}} &= \frac{\mathcal{L}(t)}{\Delta H},\quad\tau_{\text{ML}} = \frac{\pi\,\mathcal{L}^2(t)}{2\expval{H}},
\end{align}
with the Bures angle \(\mathcal{L}(t) = \arccos|F(t)|\).

Expanding to first order in \(\beta\), we write:
\begin{equation}
\mathcal{L}(t) = S_0(t) + \beta S_1(t),\quad S_0(t) = \arccos\left[e^{\alpha_0^2(\cos t-1)}\right],
\end{equation}
where the first-order correction \(S_1(t)\) is determined by Eq.~(\ref{eq:app_fidelity_expansion}).
Straightforward algebra leads to the first-order QSL corrections listed in Eqs.~(\ref{eq:sho_MT_coh_main}) and (\ref{eq:sho_ML_coh_main}).

\subsection*{Squeezed Vacuum States Under GUP Corrections}

The squeezed vacuum state remains structurally identical at first order in \(\beta\), expressed as:
\begin{equation}\label{eq:app_sq_state}
\ket{\varsigma(t)}=\sum_{n=0}^{\infty}f(2n)e^{-iE_{2n}t}\ket{2n},
\end{equation}
with coefficients:
\begin{equation}\label{eq:app_f2n}
f(2n)=\frac{\sqrt{(2n)!}\,(-\tanh r)^n}{2^n n!\sqrt{\cosh r}},
\end{equation}
unaffected at leading-order in \(\beta\).

The energy expectation \(\langle H\rangle_{\varsigma}\) and variance \(\Delta H_{\varsigma}^{2}\) in squeezed vacuum states, including first-order GUP terms, are computed as follows:
\begin{align}
\langle H\rangle_{\varsigma} 
&= \sum_{n=0}^{\infty}|f(2n)|^2 E_{2n} \nonumber\\
&= \frac{1}{2}\cosh(2r)+\frac{\beta}{16}[1+3\cosh(4r)], \label{eq:app_exp_H}\\[4pt]
\Delta H_{\varsigma}^{2} 
&= \sum_{n=0}^{\infty}|f(2n)|^2 E_{2n}^2 - \langle H\rangle_{\varsigma}^2 \nonumber\\
&= 2\,\cosh^2 r\,\sinh^2 r+\frac{3\beta}{4}\sinh(2r)\sinh(4r). \label{eq:app_var_H}
\end{align}

These results employ known sums involving hyperbolic identities:
\begin{equation}
\sum_{n=0}^{\infty}|f(2n)|^2 n = \sinh^2 r,\quad \sum_{n=0}^{\infty}|f(2n)|^2 n^2 = \frac{1}{2}\sinh^2(2r).
\end{equation}

The fidelity amplitude for initial and evolved squeezed vacuum states is given by:
\begin{equation}\label{eq:app_fidelity_sq}
\braket{\varsigma(0)}{\varsigma(t)}=\sum_{n=0}^{\infty}|f(2n)|^2 e^{-iE_{2n}t}.
\end{equation}

Expanding to first order in \(\beta\), we find explicitly:
\begin{equation}\label{eq:app_fidelity_expanded}
\begin{split}
|\braket{\varsigma(0)}{\varsigma(t)}|
={}& \frac{\sqrt{2}\,\sech r}{D^{1/4}(t)} \\[6pt]
&{}+ \beta 
   \frac{8t\,\cosh^5 r\,\sinh^2 r}
        {D^2(t)}
   \\[6pt]
&{}\frac{G(t)}
         {\bigl[1-2\cos t\,\tanh^2 r+\tanh^4 r\bigr]^{1/4}}\,.
\end{split}
\end{equation}

with the defined auxiliary functions:
\begin{align}
D(t)&=3+\cosh(4r)-2\cos t\,\sinh^2(2r),\nonumber\\[2pt]
G(t)&=-4\sin t+\sin(2t)\tanh^2 r+2\sin t\,\tanh^4 r.
\end{align}

The corresponding angular distance (Bures angle) is:
\begin{equation}
S(t)=\arccos|\braket{\varsigma(0)}{\varsigma(t)}|.
\end{equation}

To first order in \(\beta\), the angle explicitly reads:
\begin{equation}\label{eq:app_bures_angle}
S(t)=\arccos\left(\frac{\sqrt{2}\sech r}{D^{1/4}(t)}\right)+\beta\,\frac{2t\,\cosh^5 r\,\sinh^2 r\,B(t)}{A^{7/4}(t)K^{1/4}(t)Q(t)},
\end{equation}
with:
\begin{align}
A(t)&=D(t)=3+\cosh(4r)-2\cos t\,\sinh^2(2r),\nonumber\\[2pt]
K(t)&=1-2\cos t\,\tanh^2 r+\tanh^4 r,\nonumber\\[2pt]
B(t)&=G(t)=-4\sin t+\sin(2t)\tanh^2 r+2\sin t\,\tanh^4 r,\nonumber\\[2pt]
Q(t)&=\sqrt{-2+\sqrt{D(t)}}.
\end{align}

The  MT bound is explicitly derived as:
\begin{equation}
\tau_{\text{MT}}=\frac{\hbar S(t)}{\Delta H_{\varsigma}},
\end{equation}
which, to first order in \(\beta\), yields Eq.~\eqref{eq:MT_squeezed_main} in the main text.

Similarly, the ML bound is given by:
\begin{equation}
\tau_{\text{ML}}=\frac{\pi\hbar S^2(t)}{2(\langle H\rangle_{\varsigma}-E_{\min})}, \quad E_{\min}=\frac{1}{2}+\frac{3\beta}{12},
\end{equation}
explicitly expanded to first order in \(\beta\) in the main text as Eq.~\eqref{eq:ML_squeezed_main}.

\end{document}